\RequirePackage{rotating}
\RequirePackage{fix-cm}

\documentclass[epjc3]{svjour3}

\smartqed  

\usepackage{graphics}
\usepackage{graphicx}
\usepackage[usenames,dvipsnames]{color}
\usepackage{graphicx} 
\usepackage{dcolumn}
\usepackage{bm}
\usepackage{epsfig}
\usepackage{pdflscape}
\usepackage{subfigure} 
\usepackage{multirow}
\usepackage[colorlinks=true,citecolor=blue,filecolor=blue,linkcolor=blue,urlcolor=blue,pdftex]{hyperref}
\usepackage{rotating}
\usepackage{supertabular,booktabs}
\usepackage{longtable}
\usepackage{lineno}
\usepackage{array}
\usepackage{tabu}
\usepackage{longtable}
\usepackage{lscape}
\usepackage{afterpage}
\usepackage{changepage}
\usepackage{geometry}
\usepackage{verbatim}
\usepackage{cite}
\usepackage{caption}
\usepackage{amssymb}
\usepackage{amsmath}

\usepackage{fancyhdr}

\usepackage{lineno,xcolor}

\newcommand{\bologna}{Department of Physics and Astrophysics, University of Bologna and INFN-Bologna, Bologna, Italy}
\newcommand{\chicago}{Department of Physics \& Kavli Institute of Cosmological Physics, University of Chicago, Chicago, IL, USA}
\newcommand{\coimbra}{LIBPhys, Department of Physics, University of Coimbra, 3004-516 Coimbra, Portugal}
\newcommand{\columbia}{Physics Department, Columbia University, New York, NY, USA}
\newcommand{\freiburg}{Physikalisches Institut, Universit\"at Freiburg, Freiburg, Germany}
\newcommand{\lngs}{INFN-Laboratori Nazionali del Gran Sasso and Gran Sasso Science Institute, L'Aquila, Italy}
\newcommand{\mainz}{Institut f\"ur Physik \& Exzellenzcluster PRISMA, Johannes Gutenberg-Universit\"at Mainz, Mainz, Germany}
\newcommand{\heidelberg}{Max-Planck-Institut f\"ur Kernphysik, Heidelberg, Germany}
\newcommand{\munster}{Institut f\"ur Kernphysik, Westf\''alische Wilhelms-Universit\"at M\"unster, M\"unster, Germany}
\newcommand{\nikhef}{Nikhef and the University of Amsterdam, Science Park, Amsterdam, Netherlands}
\newcommand{\nyuad}{New York University Abu Dhabi, Abu Dhabi, United Arab Emirates}
\newcommand{\paris}{LPNHE, Universit\'e Pierre et Marie Curie, Universit\'e Paris Diderot, CNRS/IN2P3, Paris, France}
\newcommand{\purdue}{Department of Physics and Astronomy, Purdue University, West Lafayette, IN, USA}
\newcommand{\rpi}{Department of Physics, Applied Physics and Astronomy, Rensselaer Polytechnic Institute, Troy, NY, USA}
\newcommand{\rice}{Department of Physics and Astronomy, Rice University, Houston, TX, USA}
\newcommand{\stockholm}{Oskar Klein Centre, Department of Physics, Stockholm University, AlbaNova, Stockholm, Sweden}
\newcommand{\subatech}{SUBATECH, Ecole des Mines de Nantes, CNRS/IN2P3, Universit\'e de Nantes, Nantes, France}
\newcommand{\torino}{INFN-Torino and Osservatorio Astrofisico di Torino, Torino, Italy}
\newcommand{\ucla}{Physics \& Astronomy Department, University of California, Los Angeles, CA, USA}
\newcommand{\ucsd}{Department of Physics, University of California, San Diego, CA, USA}
\newcommand{\wis}{Department of Particle Physics and Astrophysics, Weizmann Institute of Science, Rehovot, Israel}
\newcommand{\zurich}{Physik-Institut, University of Zurich, Zurich, Switzerland}
\journalname{Eur. Phys. J. C}

\begin{document}
\sloppy 

%

\title{{Material radioassay and selection for the XENON1T dark matter experiment}}
\author{E.~Aprile\thanksref{columbia}
	    \and
	    J.~Aalbers\thanksref{nikhef}
	    \and
	    F.~Agostini\thanksref{lngs,bologna}
	    \and
	    M.~Alfonsi\thanksref{mainz}
	    \and
	    F.~D.~Amaro\thanksref{coimbra}
	    \and
	    M.~Anthony\thanksref{columbia}
	    \and
	    F.~Arneodo\thanksref{nyuad}
	    \and
	    P.~Barrow\thanksref{zurich}
	    \and
	    L.~Baudis\thanksref{zurich}
	    \and
	    B.~Bauermeister\thanksref{stockholm,mainz}
% ...
		}
\thankstext{conrad}{Wallenberg Academy Fellow}
\thankstext{lopes}{Also with Coimbra Engineering Institute, Coimbra, Portugal}
\thankstext{galloway}{email: galloway@physik.uzh.ch}
\thankstext{bern}{Also with Albert Einstein Center for Fundamental Physics, University of Bern, Bern, Switzerland}
\thankstext{xenon}{email: xenon@lngs.infn.it}
\institute{\columbia \label{columbia}
		   \and
		   \nikhef \label{nikhef}
		   \and
		   \lngs \label{lngs}
		   \and
		   \bologna \label{bologna}
		   \and
		   \mainz \label{mainz}
		   \and
		   \coimbra \label{coimbra}
		   \and
		   \nyuad \label{nyuad}
		   \and
		   \zurich \label{zurich}
		   \and
		   \stockholm \label{stockholm}
% ...
}
\date{Received: date / Revised version: date}

\maketitle

\begin{abstract}
The XENON1T dark matter experiment aims to detect Weakly Interacting Massive Particles (WIMPs) through low-energy interactions with xenon atoms. To detect such a rare event necessitates the use of radiopure materials to minimize the number of background events within the expected WIMP signal region. In this paper we report the results of an extensive material radioassay campaign for the XENON1T experiment. Using gamma-ray spectroscopy and mass spectrometry techniques, systematic measurements of trace radioactive impurities in over one hundred samples within a wide range of materials were performed. The measured activities allowed for stringent selection and placement of materials during the detector construction phase and provided the input for XENON1T detection sensitivity estimates through Monte Carlo simulations. 


\end{abstract}
%
%

\twocolumn
\section{Introduction}
\label{sec:intro}

Observations at astronomical and cosmological scales indicate that a majority of the matter content of our Universe is in the form of non-relativistic, long-lived, and non-luminous dark matter~\cite{PDC, GalaxyFormation, LargeScaleStructure, Clowe}. Extensions of the Standard Model favour a candidate for dark matter in the form of a Weakly Interacting Massive Particle (WIMP)~\cite{Steigman, Jungman}. Its interaction with normal matter can be probed directly via elastic scattering off target nuclei, thus motivating searches through direct detection \cite{Goodman:1984dc}. The XENON collaboration has constructed and commissioned the first ton-scale liquid-xenon dark matter detector, aiming to observe primarily low-energy nuclear recoils of WIMPs with unprecedented sensitivity.

XENON1T, a dual-phase time projection chamber (TPC)~\cite{Bolozdynya, Xe1T_TPC}, was designed to improve the sensitivity of its predecessor, XENON100~\cite{xenon_xe100-SI, xenon_xe100-SD}, by two orders of magnitude for the spin-independent WIMP-nucleon interaction cross section. The increased sensitivity is achieved through reducing the background and increasing the target mass, i.e. the amount of liquid xenon (LXe) enclosed by the TPC, by a factor of 32, allowing for a sensitive volume, after fiducialization, of $\sim$1 ton. In rare-event searches, understanding and minimizing background events that occur within the sensitive volume of the detector is of utmost importance. This necessitates the use of construction materials with low intrinsic radioactivity, passive and active detector shielding, and sophisticated analysis techniques in order to prevent background events within the parameter space where a WIMP signal is expected. 
 
The XENON1T radioassay program addresses backgrounds that come from radioactive impurities within detector construction materials. Radioassay of candidate materials provides information about the type and amount of expected emissions, thus allowing for selection and strategic placement of the most radiopure materials within the detector. The measured results provide the material-induced radiogenic component to the overall background model of XENON1T. Through Monte Carlo simulations using the activities from each component, accurate predictions of the detector sensitivity were performed~\cite{xenon_xe1t-sensitivity}.

Here we present an overview of the screening and material selection process for XENON1T. Section~\ref{sec:bkgrd} describes expected background sources and reduction methods. The techniques and instruments used to identify and estimate radioimpurities of each sample are detailed in Section~\ref{sec:techniques}. Section~\ref{sec:results} describes the various materials and components that were screened with respect to the decay chains and isotopes that are of greatest concern. For each relevant decay chain and single-line emission, the radioassay results are presented in Table~\ref{table:Samples}. We summarize the results in Section~\ref{sec:impact} with a discussion of the impact on the XENON1T sensitivity with respect to the materials measured in this study.

\section{Background expectation and minimization}
\label{sec:bkgrd}

Particle interactions with either atomic electrons or nuclei of the xenon target result in electronic recoil (ER) events or nuclear recoil (NR) events, respectively. The nuclear recoil background, predominantly from neutrons, is the most dangerous, as the signature of a WIMP is a single-scatter, NR event. Background discrimination and rejection techniques include removing multiple-scatter events, fiducializing the target volume through event vertex reconstruction, and exploiting the difference in energy loss per unit track length between ER and NR events~\cite{xe100analysis}. However, ER events that occur within the fiducial volume can leak into the NR region of the WIMP discrimination parameter space as well as obscure other rare event searches that are otherwise possible in the ER channel. The aim, therefore, is to mitigate sources of both types of backgrounds and to accurately estimate the number of expected background events within the WIMP search region. 

External background from cosmic rays, i.e.~hadronic components and muon-induced neutrons, is suppressed by operating the detector at an average depth of 3600 meters water equivalent in the Laboratori Nazionali del Gran Sasso (LNGS), thus reducing the muon flux by a factor of 10$^{6}$ relative to a flat overburden~\cite{LNGS_muonflux}. A water shield instrumented with veto PMTs surrounds the detector by at least 4 meters on all sides to provide passive shielding and to reject coincident events detected via Cherenkov radiation~\cite{xenon_xe1t-MuonVeto}. Solar neutrinos are another potential source of external background, both ER and NR, the latter from coherent neutrino-nucleus scattering. 

Sources of ER background that are intrinsic to the xenon target, e.g. the beta emitter $^{85}$Kr, and the double-beta emitter $^{136}$Xe, are expected to be uniformly distributed throughout the xenon, thus cannot be reduced through fiducialization. However $^{85}$Kr can be significantly reduced through distillation from $^{\mathrm{nat}}$Kr to $<0.2$ ppt~\cite{Kr_column}, and $^{136}$Xe, comprising 8.9$\%$ of natural xenon, has a subdominant contribution of $<2\%$ to the total ER background~\cite{xenon_xe1t-sensitivity}. Although not natively intrinsic to the scintillator, the noble gas $^{222}$Rn (T$_{1/2}=$3.8 d), originating from the long-lived $^{226}$Ra (T$_{1/2}=$1600 yr), mixes with the xenon and becomes homogeneously distributed within the target. Beta decays of its daughter isotopes are the dominant source of ER background. Moreover, $^{214}$Pb and the daughter isotopes from its decay to ground state adhere to material surfaces (plate-out) and can lead to \mbox{($\alpha$, n)} reactions within the target volume. Because of plate-out effects from both parent and daughter isotopes of $^{222}$Rn, the level of contamination for this isotope is determined by directly measuring its emanation from construction materials. This technique will be described in a separate publication~\cite{RnEmanation}. Additionally, a significant reduction in radon by online purification has recently been demonstrated by the XENON collaboration through the use of a cryogenic distillation technique~\cite{RnDistillation}.

The radioassay program described in this paper targets the background from radionuclei present as residual traces in the detector components. The most common radioactive contaminants are long-lived (T$_{1/2} >$ 1 yr) primordial radionuclei within the $^{238}$U and $^{232}$Th decay chains and the single isotope $^{40}$K. The latter isotope as well as several isotopes within the primordial chains decay via high-energy gamma emissions that cannot be completely removed through fiducialization. Additionally, several isotopes belonging to these chains release neutrons either directly through spontaneous fission of heavy nuclei or indirectly via \mbox{($\alpha$, n)} reactions following alpha decays within the chains. 

In addition to primordial radioisotopes, anthropogenic radioisotopes, such as $^{137}$Cs and $^{110\mathrm{m}}$Ag, can be found in some detector materials. These isotopes are either manufactured for medical or industrial use or are generated from nuclear power plant waste, nuclear accidents, or military testing. Cosmogenic isotopes, such as $^{54}$Mn, $^{46}$Sc, and $^{56-58}$Co, can be found mainly in metal components as a result of activation from exposure to cosmic rays~\cite{cosmogenic}. An additional common radionuclide is $^{60}$Co, which is primarily anthropogenic in origin in stainless steel and cosmogenically induced in copper. Most of the listed radionuclei, including many of the isotopes within the primordial decay chains, can be detected with high sensitivity by the XENON1T radioassay techniques described in Section~\ref{sec:techniques}.

\section{Techniques and measurements}
\label{sec:techniques}

To determine the amount and isotopic composition of radionuclides present in the XENON1T materials, gamma-ray spectroscopy and mass spectrometry methods were used. The former provides a non-destructive technique sensitive to almost every relevant gamma emitter and allows to detect a break of secular equilibrium within the primordial decay chains. To reach the detection sensitivity required by current dark matter search experiments, e.g. at or below the $\mu$Bq/kg level for some materials, large sample masses ($\sim10$$-$$20\,$kg) and long counting times ($\sim15$$-$$20\,$d) are usually necessary. This is particularly the case for low-energy gamma emitters due to self-absorption within the sample volume.

Mass spectrometry, in particular, Inductively Coupled Plasma Mass Spectrometry (ICP-MS) and Glow Discharge Mass Spectrometry (GDMS) were used to assess the composition of a given sample through separation and measurement of individual isotopes. This is particularly useful in determining the amount of $^{238}$U and $^{232}$Th within materials. Because ICP-MS and GDMS require just a few grams of sample material and short measurement times, they are also used when the mass of the sample is too small or the available measurement time too short to achieve the desired sensitivity in an HPGe spectrometer.

\subsection{Germanium spectrometers}
\label{sec:Ge}

The XENON collaboration utilizes several of the world's most sensitive germanium spectrometers, the Gator~\cite{gator} detector and the four GeMPI detectors~\cite{gempi}, that are located in ultra-low background facilities at LNGS at the same depth as the XENON1T detector. These spectrometers have an excellent energy resolution over the energy range of interest (\mbox{$\sim50 - 2650\,$ keV} with, e.~g. \mbox{$<3$ keV} FWHM at \mbox{1332 keV}) and can reach sensitivities down to the $\mu$Bq/kg level. All detectors are p-type, intrinsically pure germanium crystals in a coaxial configuration, with masses between \mbox{2.2$-$2.3 kg} and enclosed in a low-background cryostat housing. The sensitive region of the cryostat protrudes into an inner chamber made of electro-refined, Oxygen-Free High-Conductivity (OFHC) copper, with a material sample capacity of several liters in volume. The inner chamber is constantly purged with pure nitrogen to suppress the influx of radon. The copper is surrounded by a \mbox{15$-$25} cm thick lead shield, where the innermost layer of \mbox{2$-$5 cm} has a low level of $^{210}$Pb contamination. Radon-free nitrogen-flushed glove boxes are located on top of each detector.

The radioassay program used three additional spectrometers, Corrado, Bruno, and GIOVE~\cite{giove}, that are operated underground in the Low-Level Laboratory at Max Planck Institut f{\"u}r Kernphysik (MPIK) in Heidelberg. The laboratory has an overburden of 15 meters water equivalent that reduces the muon flux by a factor of 2$-$3 and the hadronic background component by a factor of 1000 as compared to sea level. The spectrometers are surrounded by copper and lead shielding. Additionally, an active muon veto and polyethylene to moderate neutrons surround the Giove detector. These three detectors are p-type germanium crystals with masses between 0.9$-$1.8 kg that can reach sensitivities between 0.1$-$1 mBq/kg. 

Given the higher background and lower detection sensitivity with respect to the spectrometers operated at LNGS, the MPIK detectors were mostly employed for radioassay of components that are far from the sensitive volume of the XENON1T TPC, such as the tank for the water shield as well as the support structures and calibration systems within the shield. Most of the materials from components closest to the active volume of the TPC were screened with the GeMPI or Gator detectors at LNGS. For several smaller samples, additional detectors at the LNGS underground low-background facility STELLA (SubTerranean Low Level Assay) were used~\cite{GePaolo}.

Samples were cleaned using the same techniques as in the final detector construction when possible. Otherwise, the standard procedure was to clean each sample with a mild acid soap (Elma EC70), followed by rinsing with deionized water and immersion in high-purity ethanol \mbox{($>95$\%)}. Each step utilized an ultrasonic bath for 20 minutes. Samples where acid soaps or immersion in liquids should be avoided, such as photomultiplier tubes and cables, were cleaned by wiping the surface with ethanol. During transport of the samples to the detector glovebox, they were either enclosed in clean plastic bags or wrapped in plastic foils in order to prevent the plate-out of $^{222}$Rn daughters from the environment. The samples were then stored in an outer glovebox of the detector prior to their measurement in order to let the radon and its daughters decay.

For every measured sample, a Monte Carlo simulation based on the GEANT4 toolkit~\cite{Agostinelli} was used to calculate the detection efficiencies for each emitted gamma line. The efficiencies were used in combination with the sample mass, measurement time, and branching ratio of each characteristic \mbox{gamma-ray} line to determine the specific activities or detection upper limits of each radioisotope. Further details on analysis procedures for the HPGe detectors can be found in~\cite{gempi, gator}.

\subsection{Inductively Coupled Plasma Mass Spectrometry}

Inductively Coupled Plasma Mass Spectrometry is one of the most sensitive analytical techniques for trace element analysis. The intrinsic radioactivity of materials can be deduced by measuring the concentration of long-lived radionuclides. The sample (fractions of a gram are enough for a measurement) is introduced as an aqueous solution through a peristaltic pump, nebulized in a spray chamber, then atomized and ionized in a plasma. The ions are extracted into an ultra-high vacuum system and separated according to their mass-to-charge ratio by the mass analyzer. The concentration of the ions is calculated based on the response of reference standard solutions. Depending on the nature of the sample, sensitivities on the order of 10$^{-11}$ to 10$^{-13}$ g/g for $^{238}$U and $^{232}$Th and 10$^{-7}$ to 10$^{-8}$ g/g for $^{39}$K can be reached. This corresponds to activities of 1$-$100 $\mu$Bq/kg and a few mBq/kg, respectively. The uncertainty of measurement is between 20$-30\%$ and accounts for several factors, such as the instrumental precision, the single replicate measurement and the single-point calibration curve.

For this radioassay campaign, measurements were performed with a 7500a ICP-MS from Agilent Technologies and an Element II HR-ICP-MS from ThermoFisher Scientific. Both instruments are located in a ISO6 clean room at the Chemistry Laboratory of LNGS.

\subsection{Glow-discharge mass spectrometry}

The Glow Discharge Mass Spectrometry measurements for XENON1T were performed at EAG Laboratories~\cite{EAG}. Rather than being introduced as an aerosol as in ICP-MS, a negative bias is applied to the solid sample material while exposed to an argon-based plasma in order to induce sputtering via ion-target collisions. Once the material is sputtered into the plasma and subsequently ionized, an ion beam is extracted and focused through a high resolution mass spectrometer. Ions are separated according to their \mbox{mass-to-charge} ratio. Sensitivity of \mbox{sub-ppb} level or 10$^{-10}$ g/g ($\sim$1 mBq/kg) can be reached with an uncertainty between 20$-30\%$. Electrical conductivity of the sample is needed for stable and reproducible glow formation, thus the reliability and sensitivity of GDMS may vary depending upon properties of the target material. As with ICP-MS, GDMS is particularly useful in determining the $^{238}$U and $^{232}$Th concentrations. Although ICP-MS provides a better sensitivity than GDMS, the choice to use GDMS was primarily due to the availability and location of the measurement facilities.\\


\section{Radioassay results}
\label{sec:results}

\begin{figure}[]
\begin{center}
\includegraphics*[width=0.9\linewidth]{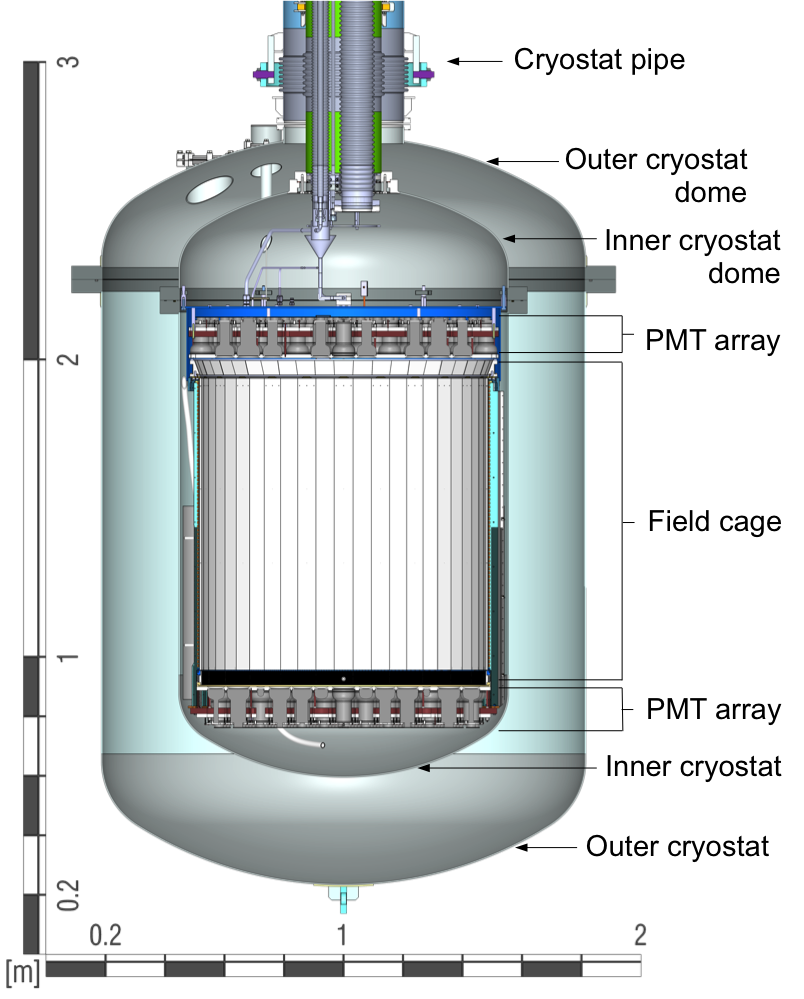}
\end{center}
\caption
{\label{fig:TPC_section}The XENON1T TPC with cryostat, section view, subgroups are indicated with reference to the ``XENON1T Use" column of Table~\ref{table:Samples}.}
\setlength{\belowcaptionskip}{0pt}
\end{figure}

\begin{figure}[]
\begin{center}
\includegraphics*[width=.7\linewidth]{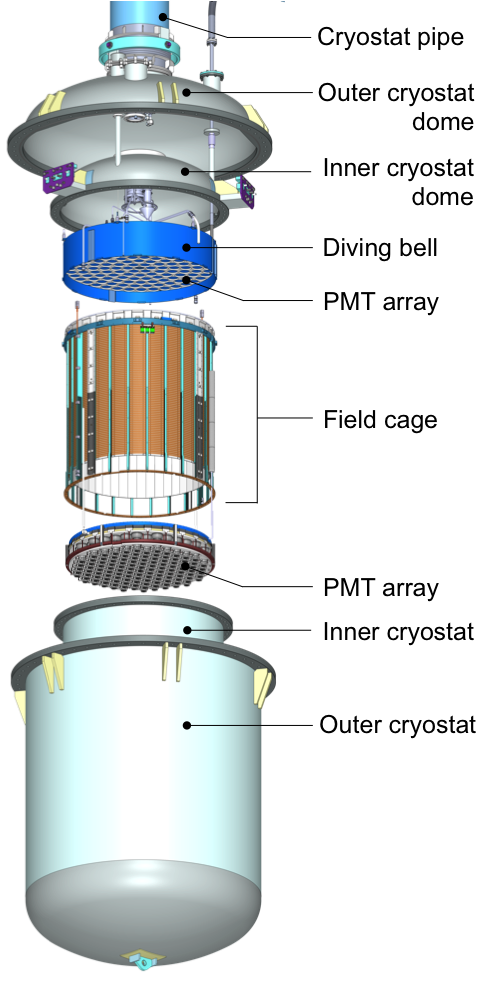}
\end{center}
\caption
{\label{fig:TPC_exploded}The XENON1T TPC with cryostat, subgroups are indicated with reference to the ``XENON1T Use" column of Table~\ref{table:Samples}.}
\setlength{\belowcaptionskip}{0pt}
\end{figure}

Results obtained through the radioassay program are shown in Table~\ref{table:Samples}. Throughout the text, the samples are identified by their unique item numbers (``$\#$"). The detector is shown in Figs. \ref{fig:TPC_section} and \ref{fig:TPC_exploded} to introduce the most relevant subgroups and components. These are given in the ``XENON1T Use" column of Table~\ref{table:Samples} in the case where the material or component was chosen for detector construction. 

Supplier information is provided where applicable. For measurements conducted with the HPGe spectrometers, the sample mass and measurement duration are noted. Uncertainties, including both statistical and systematic, the latter primarily from efficiency simulations, are given in parentheses as $\pm$ 1$\sigma$ of detected activities or at 95$\%$ confidence level for upper limits. Unless otherwise specified, the uncertainties of ICP-MS and GDMS measurements are 30$\%$ and are primarily systematic, as described in Section 3.2.

The upper part of the $^{232}$Th decay chain is measured directly by mass spectrometry methods but is only detectable from gamma-ray spectroscopy following the $^{228}$Ra decay \mbox{(T$_{1/2}= 5.7$ yr)}. Rather than assuming secular equilibrium in the upper part of this chain, the two results are presented together with an indication of which part of the chain was measured. It is worth noting that, with one exception (copper $\#$4), all samples for which both $^{232}$Th with ICP$-$MS and $^{228}$Ra with HPGe spectroscopy were measured show consistent results, thus indicating no break in secular equilibrium at $^{228}$Ra for these samples.

In addition to the decay chain and radioisotope activities listed in Table~\ref{table:Samples}, Table~\ref{table:metalother} shows results from cosmogenic radionuclei with \mbox{short-to-moderate} \mbox{half-lives} (T$_{1/2} <$ 1 yr), as detected with HPGe spectrometers. 

All results reported here will be made available via the radioassay community database at http://www.radiopurity.org.~\cite{persephone}. Further details on many of the XENON1T samples and their measurements can be found in~\cite{Piastra}.

\subsection{Metal samples}
\label{sec:metal_samples}

Commercially available Oxygen-Free High Conductivity or Oxygen-Free Electrolytic copper (OFHC or OFE copper) from primarily two different distributors was used for several major components of the TPC: the 74 field-shaping rings that surround the TPC ($\#$1), the top and bottom PMT array support structures ($\#$2, $\#$3), and the bottom structural ring of the field cage ($\#$4), comprising $\sim$190 kg of the detector mass. Copper is intrinsically radiopure, with detected activities of the natural decay chains at the ppt level (see Table~\ref{table:Samples}). One can see that the $^{60}$Co activity from cosmogenic activation varies from batch to batch, depending on the storage and shipment of the material~\cite{cosmometal}. Because of its relative purity, copper was used as a substitute for stainless steel wherever possible. A sample of copper plated with 2 $\mu$m thick gold ($\#$6) was considered for the field-shaping rings to minimize radon emanation, however the samples showed significantly higher $^{40}$K activity that was most likely introduced as part of the electrochemical plating process. 

The radiopurity of stainless steel can vary between batches, depending upon the source of the raw material, the method of heating and forming the material, as well as the location and duration of storage of the metal (cosmogenic activation). In order to minimize emissions from stainless steel components near the sensitive volume, the cleanest batches of available material were required. 
Therefore many batches of 304 and 316 stainless steel from six different manufacturers (17 samples in total) were screened for radiopurity. The samples originated from different melts and consisted of varying thicknesses. The NIRONIT Edelstahlhandel GmbH $\&$ Co. samples ($\#$8$-$10) that were particularly low in $^{226}$Ra, $^{232}$Th, and $^{60}$Co were selected for production of various TPC components and the cryostat vessels. Materials for components that are in direct contact with the liquid xenon, such as the cryostat pipe ($\#$11), were selected for low $^{226}$Ra contamination in order to minimize emanation of $^{222}$Ra that can mix with the xenon and end up in the fiducial volume. 

In many of the stainless steel batches, a depletion in $^{226}$Ra with respect to the upper half of the $^{238}$U chain is observed, thus indicating a clear break in secular equilibrium that most likely occurred during processing of the raw materials. In particular, a disequilibrium of more than a factor of 10 can be seen for items $\#$17, $\#$23, and $\#$30. Additionally, a break between the upper and lower parts of the $^{232}$Th chain is observed in some of the HPGe measurements ($\#$12$-$14, $\#$17, $\#$25, $\#$30). Because these results were obtained through HPGe spectroscopy, the $^{232}$Th activity was not measured directly, thus the break in this chain is a possible indication of depletion in $^{228}$Ra. 

The induced background from the XENON1T structural components, such as the water tank and outer support structures ($\#$24$-$27, in addition to many screened samples not listed), was shown to be negligible in the Monte Carlo simulations due to their distance from the sensitive volume~\cite{xenon_xe1t-sensitivity}. The screened stainless steel hardware ($\#$28, $\#$29) used for critical internal components, such as for the resistor chain and electrode fasteners, also had a negligible background contribution as the total mass used in the final construction was less than 1 kg. 

Titanium was considered as a potential cryostat material because of its high tensile strength as compared to copper and potentially lower radioactivity as compared to stainless steel. It has previously been used in the LUX experiment~\cite{LUX_titanium} and investigated for use in the upcoming LZ experiment~\cite{LZ_titanium}. Three different grades of titanium from four different suppliers were measured. The measured contamination of the titanium samples ($\#$32$-$40) showed roughly a factor of 10 higher activity in the uranium chain as compared to the stainless steel used for the cryostat ($\#$9). The other difference in contamination between the two material types was with respect to $^{60}$Co, which is subdominant in titanium but of concern in stainless steel, and $^{46}$Sc, a prominent cosmogenic isotope in titanium, as shown in Tables~\ref{table:Samples} and~\ref{table:metalother}, respectively. Additionally, the lower mechanical strength of titanium as compared to stainless steel would have required a thicker cryostat. When taking this into account in the Monte Carlo simulations, the neutron background from a titanium cryostat was considerably higher than for its stainless steel counterpart, therefore the latter material was chosen to construct the XENON1T cryostat.

\subsection{Plastic samples}
\label{sec:plastic_samples}

Due to its good VUV reflectivity ($>$95$\%$), a dielectric constant similar to liquid xenon, low-outgassing properties as compared to other plastics, and machinability, polytetrafluoroethylene (PTFE) is the material of choice for reflective surfaces within the field cage. Because it directly encloses the LXe sensitive volume, its radioactive content must be sufficiently low and also precisely measured to achieve an accurate background estimate. All of the PTFE samples were measured using ICP-MS for better quantification of the primordial chain progenitor isotopes and to complement the HPGe measurements where available, showing levels typically at the tens of ppt or $\mu$Bq/kg level ($\#$46$-$50). PTFE doped with 15$\%$ quartz to increase the reflectivity ($\#$51) was also measured, however showed gross contamination in all of the natural chains as seen in Table~\ref{table:Samples}. The primordial chains, due to alpha decays, are of particular concern for PTFE, as neutrons can be generated in the material via $^{19}$F($\alpha$, n) reactions~\cite{Norman}. Thus efforts were also made to minimize the total amount of this material used in construction. 

Polyamide-imide (PAI, in this case Torlon 4203L) was investigated for use as an insulating, structural material as it has a high dielectric constant, good mechanical strength, low thermal contraction, and allows for high-precision machining. Radioassay results ($\#$53, $\#$54) showed activities from the primordial chains to be a factor of 10 to 100 higher than its structural counterpart, PTFE. However, due to the absence of fluorine, neutron emission via ($\alpha$, n) is not an issue with PAI. It was used for small but critical components, e.g. as insulating spacers. 

Commercially available PEEK (polyether ether ketone) screws were used at locations inside the TPC that required a high dielectric constant but with limited load-bearing requirements. Only one PEEK sample was measured ($\#$52), yielding results on the order of 1$-$10 mBq/kg, comparable to that of PAI.

For all of the plastic samples, no clear break in secular equilibrium is observed in the primordial decay chains. However the case of equilibrium is inconclusive, as only upper limits were measured for most samples. One exception is the PTFE doped with quartz ($\#$51), that shows a clear break in the $^{232}$Th chain.

\subsection{Photomultiplier tubes and related components}
\label{sec:pmts}
 
The radioactive budget of the Hamamatsu R11410 3-inch diameter photomultiplier tubes was initially estimated through screening of the raw materials used in fabrication. Subsequently, several versions of PMTs were produced and screened with the goal of minimizing the total radioactivity of the tube to arrive at the final version, R11410-21. Of this version, the averaged activities of 320 PMTs measured with Gator and 40 PMTs measured with GeMPI I are reported in Table~\ref{table:Samples} ($\#$69, $\#$70). Where only an upper limit was found, no entry is provided. Further details on the specific material contributions and the development of these low-background photomultipliers are given in~\cite{LowRadPMTs}.

Several samples of cables for the PMTs were screened to find clean batches. The detected activities for the signal and high-voltage cabling ($\#$55$-$56, and $\#$57$-$58, respectively) that were selected for final construction were typically at the tens or lower mBq/kg level, with the exception of the considerably higher presence of $^{40}$K, particularly in the high-voltage (kapton) cables. The remaining PTFE ($\#$59$-$61) and kapton coaxial and flat cables ($\#$62$-$64) were not used due to higher levels from the primordial decay chains. 

The connectors for the PMT signal and high-voltage cables, respectively, consisted of male/female pairs of micro-miniature coaxial (MMCX) connectors made from a copper-zinc alloy ($\#$66) and of subminiature-D (D-sub) pins made from either a copper-beryllium alloy ($\#$65, $\#$68) or a gold-plated copper alloy ($\#$67). Due to the minimal total mass and the locations of the connector assemblies relative to the sensitive volume (in the cryostat pipe and on top of the diving bell), their radioimpurities are considered to have a negligible contribution to the overall background budget. Therefore all of the screened batches were used in the final construction. Additionally, a measurement of a representative sample of the high-voltage connectors mounted in custom-made PTFE holders, as produced for the final assembly, was performed with a new HPGe spectrometer, GeMSE, and showed consistent results~\cite{GeMSE}.

Connected directly to the base of each PMT is a voltage divider network that consists of a Printed Circuit Board (PCB, $\#$94) with sockets ($\#$93), solder ($\#$99), resistors ($\#81-84$), and capacitors ($\#$86, $\#$88). Several batches of the same types of components were screened, as there was some variation among batches and with respect to different PCB materials. The final PCBs assembled with components (referred to as the PMT base in Table~\ref{table:Samples}) used the cleanest components where possible and then screened with an HPGe spectrometer ($\#$100). The activity per assembled base was measured to be about a factor of 10 lower than the activity from the PMT itself.

Several of the components for the PMTs show a clear break in secular equilibrium in the $^{238}$U chain indicating a depletion of $^{226}$Ra, particularly the connectors ($\#$66), the sockets ($\#$93), many of the resistors, and, consequently, the assembled bases ($\#$100). 

\subsection{Other samples}
\label{sec:other_samples}

Several components that were composites of different materials, such as insulated conductors for electrode high voltage (a copper rod inserted into a PTFE insulator, $\#$103) and capacitive sensors to measure the LXe level (``levelmeters", $\#$101$-$102) were screened post-fabrication and showed acceptable activities. The remaining components listed under ``Miscellaneous'' showed high levels in the primordial decay chains. However, these components are used in the calibration or leveling systems that are located within or outside of the Cherenkov detector and quite far from the TPC sensitive volume, therefore have negligible contributions to the background.


\subsection{Summary of material placement}
\label{sec:instrument}

\begin{figure}[]
\begin{center}
\includegraphics*[width=0.7\linewidth]{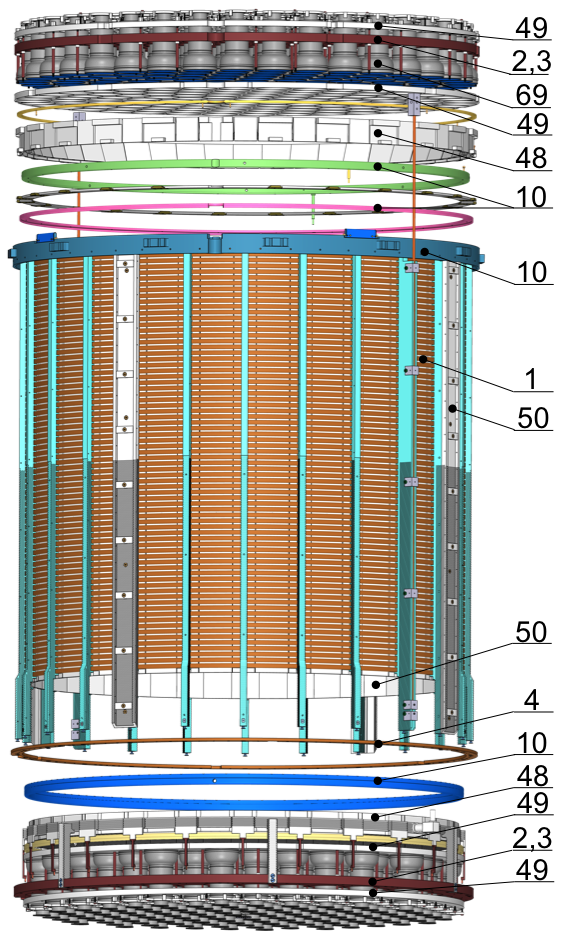}
\end{center}
\caption
{\label{fig:TPC_items}The XENON1T TPC with material item numbers as given in Table~\ref{table:Samples}.}
\setlength{\belowcaptionskip}{0pt}
\end{figure}

The contribution from each material to the background depends upon its total mass and proximity to the sensitive volume as well as its type and energy of emission. The locations of screened materials used for the major components of the XENON1T TPC are indicated in Fig~\ref{fig:TPC_items} by item number. The radioassay results from Table~\ref{table:Samples} in combination with the material distribution within the instrument informed the XENON1T background predictions, as described in~\cite{xenon_xe1t-sensitivity}. 

The field cage of the XENON1T TPC consists of PTFE reflector panels and support pillars ($\#$50), the latter hold and maintain separation between the 74 high-purity copper field-shaping rings ($\#$1). The bottom ends of the PTFE pillars are mounted to a copper ring ($\#$4) and are supported on the top by a stainless steel ring ($\#$10). Bottom and top arrays of photomultiplier tubes ($\#$69, $\#$70, $\#$100) face the target liquid-xenon volume enclosed by the field cage. The bottom array consists of a copper support plate ($\#$2, $\#$3) with a PTFE layer underneath ($\#$49) for electrical insulation and a polished PTFE surface ($\#$49) at a stand-off distance above the Cu plate in order to reflect the VUV light from the surfaces surrounding the PMT photocathodes. The top array consists of the same layers as the bottom array, mounted upside-down inside of the stainless steel diving bell that controls the LXe level ($\#$10, shown in Fig~\ref{fig:TPC_exploded}). In front of the photocathode surfaces of each PMT array are stainless steel screening electrodes ($\#$10, not indicated in Fig~\ref{fig:TPC_items}) to protect the PMTs from the field cage high voltage, small PTFE reflectors ($\#$48), and the three electrodes ($\#$10) that provide the electric field across the TPC (cathode below the target, gate and anode electrodes above the target). 

Components not shown in Fig~\ref{fig:TPC_items} include small parts such as the 5 G$\mathrm{\Omega}$ resistors ($\#$71) that connect neighbouring copper field-shaping rings, PMT cabling and connectors ($\#$55$-$58, $\#$65$-$68), and small copper ($\#$1), PEEK ($\#$52), and stainless steel screws ($\#$8, $\#$28, $\#$29) that were used throughout the TPC. Also not shown are components mounted onto or near the top stainless steel ring such as PAI ($\#$53, $\#$54) and PTFE ($\#$50, $\#$103) insulating spacers, and the levelmeters ($\#$102) which are used to precisely measure the vertical position of the xenon liquid/gas interface. 

Other TPC components (not shown in Fig~\ref{fig:TPC_items}) are two long levelmeters ($\#$101) which are used during LXe filling and a stainless steel with polyethylene high-voltage feedthrough (made from $\#$8, $\#$108) inside of a PTFE insulator ($\#$47) that span the length of the field cage. The PMT signal and high-voltage cables ($\#$55, $\#$57) extend from the bottom PMT array along the length of the field cage and from the top PMT array inside the diving bell. The cables are then routed over the diving bell and connect to the cables ($\#$56, $\#$58, connected by $\#$65$-$68) that arrive from the data acquisition room via the cryostat pipe ($\#$11, partially shown in Figs~\ref{fig:TPC_section} and~\ref{fig:TPC_exploded}). 

The cryostat ($\#$9), shown in Figs~\ref{fig:TPC_section} and~\ref{fig:TPC_exploded}, consists of an inner stainless steel vessel that encloses the TPC and liquid xenon, nested inside an outer vessel that is evacuated for thermal insulation. The cryostat vessels and their domes ($\#$8, $\#$9) are covered by mylar insulation ($\#$41) to reduce heat losses. Not shown are components outside of the cryostat, such as the calibration systems ($\#$104, $\#$105, $\#$107) and stainless steel support structures ($\#$25$-$27) within the water shield, and the 10 meter high, 9.6 meter diameter stainless steel tank that contains the water shield ($\#$24).

\section{Discussion and impact on the XENON1T background}\label{sec:impact}

The results from the radioassay campaign were used as source terms in the detector Monte Carlo simulations. The detected radioactive isotopes and decay chains were uniformly distributed within each component of the mass model according to their measured radioactivities. Each background source, before ER/NR descrimination, is given in terms of an event rate over a \mbox{1 ton} super-ellipsoid fiducial volume with respect to the energy region of interest (ROI). As electronic recoils and nuclear recoils induce a different response in liquid xenon, nuclear recoils in the (4, 50) keV$_{\mathrm{nr}}$ interval yield the same signal intensities from scintillation as ER events in the (1, 12) keV$_{\mathrm{ee}}$ (electron equivalent) energy ROI. The simulation and analysis details are given in~\cite{xenon_xe1t-sensitivity}. 

\begin{figure}[h]

\begin{center}
\includegraphics*[width=1.0\linewidth]{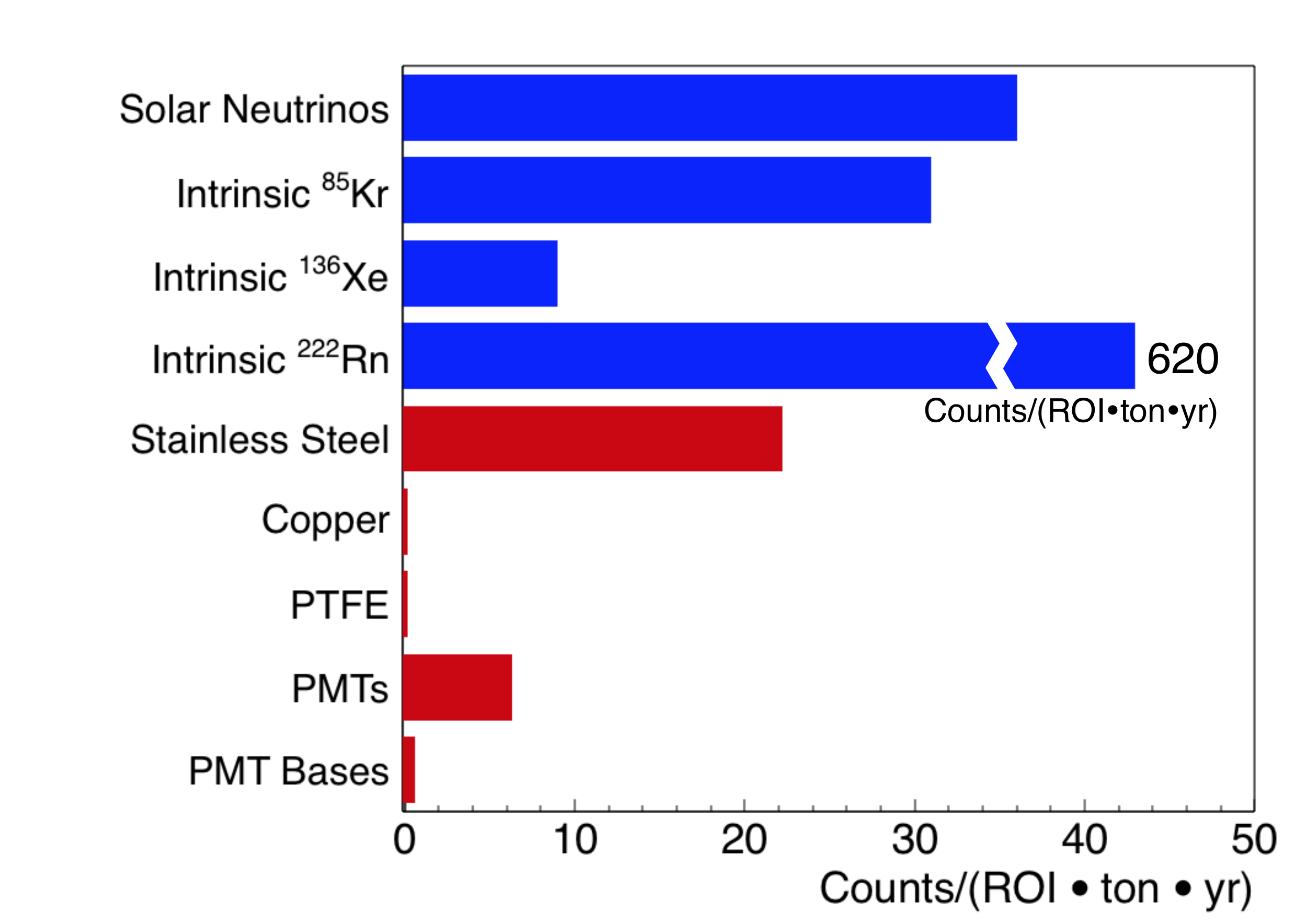}
\includegraphics*[width=1.0\linewidth]{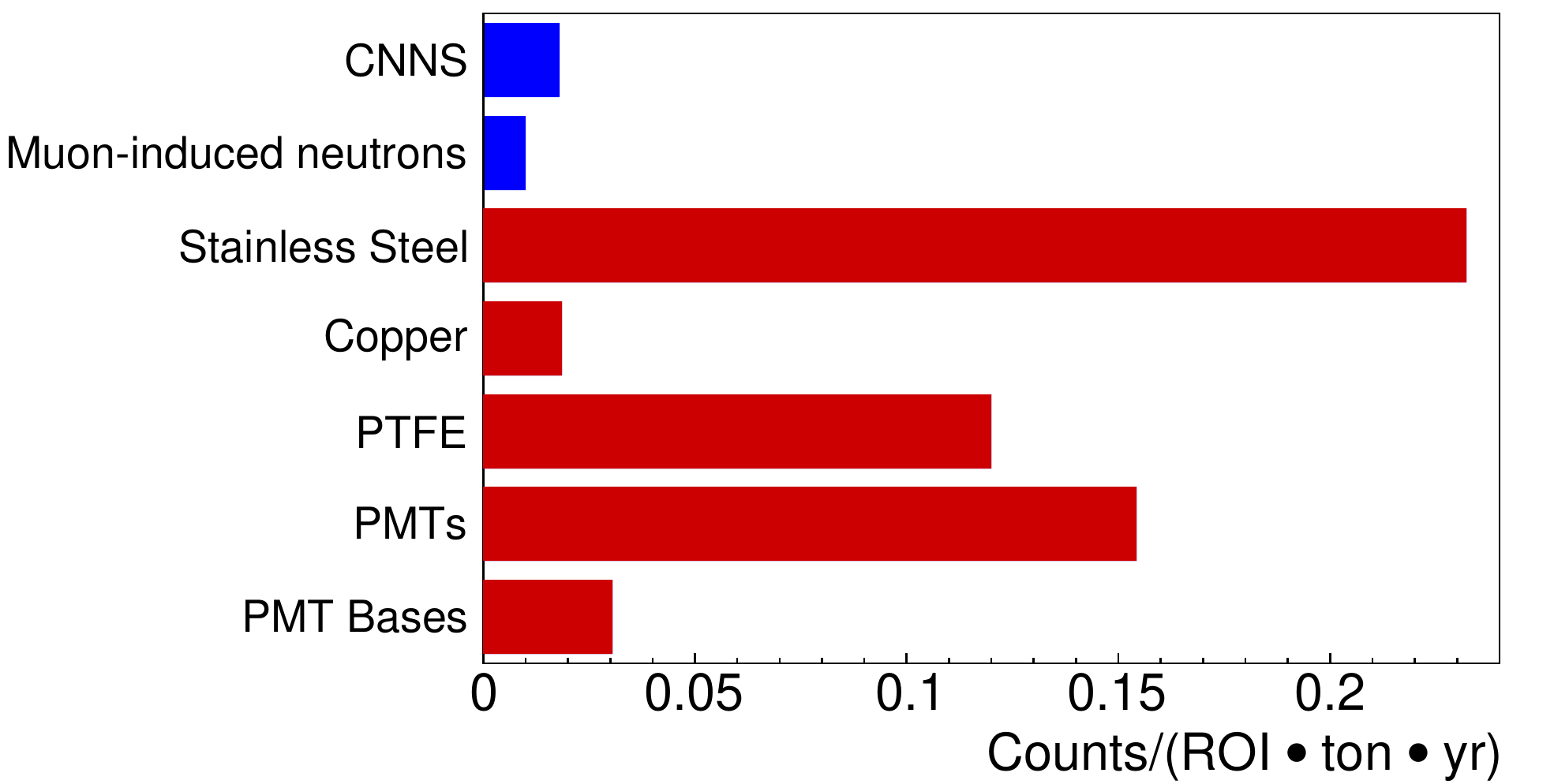}
\end{center}

\caption
{\label{fig:counts_All}Electronic recoil (top) and nuclear recoil (bottom) background contributions from materials (red) and from intrinsic and external sources (blue). The number of events per year in a 1$-$ton fiducial target is shown in the electron equivalent (1, 12) keV$_{\mathrm{ee}}$ region of interest for electronic recoil events, corresponding to a nuclear recoil energy interval of (4, 50) keV$_{\mathrm{nr}}$.}
\end{figure}

Figure \ref{fig:counts_All}, top, shows the relative expected contributions to the total ER background events for external background sources (i.e.~solar neutrinos), sources of intrinsic backgrounds ($^{136}$Xe, $^{85}$Kr, and $^{222}$Rn), and for each of the main XENON1T components. Thanks to the material selection campaign described in this work, the material-induced gamma-ray background is negligible within the (1, 12) keV$_{\mathrm{ee}}$ WIMP search region compared to the contribution from $^{222}$Rn emanation. The dominant intrinsic $^{222}$Rn contamination was estimated to be 10 $\mu$Bq/kg in the liquid xenon target, however this can be further reduced through online purification~\cite{RnDistillation}. A more detailed comparison with respect to the energy and select fiducial volumes can be found in~\cite{xenon_xe1t-sensitivity}.


The expected contributions to the nuclear recoil background in XENON1T are shown in Fig. \ref{fig:counts_All}, bottom. Most of the NR background comes from materials, as there are no significant intrinsic sources. Considering materials only, the stainless steel components (cryostat, TPC) are the dominant source, in total contributing 40$\%$. The PMTs contribute 28$\%$, primarily due to the high concentration of $^{238}$U and $^{232}$Th and their daughter isotopes in the ceramic stem of each PMT. Because of the proximity of the PTFE reflectors to the sensitive volume, the presence of heavier nuclei and their daughters contribute 22$\%$ by the mechanisms described in Section~\ref{sec:plastic_samples}. Coherent neutrino-nucleus scattering (CNNS) is subdominant, with a contribution similar to the TPC copper ($\sim3\%$ of total). The muon-induced nuclear recoil background is also subdominant due to effective coincidence-tagging with the Cherenkov muon-veto detector~\cite{xenon_xe1t-MuonVeto}. 

After conversion into observable signals, ER/NR discrimination was applied to all background events. Assuming an ER rejection efficiency of \mbox{99.75$\%$} at an NR acceptance of 40$\%$, the total expected NR background in XENON1T for a 1 ton~$\times$~2 year exposure is expected to be \textless1 event in the (4, 50)\,keV$_{\mathrm{nr}}$ energy range. This corresponds to a best sensitivity to the spin-independent WIMP-nucleon cross section of \mbox{$\sigma_\mathrm{SI} \lesssim 10^{-47}~\mathrm{cm}^{2}$} at a WIMP mass of \mbox{m$_{\chi}$= 50 GeV/c $^{2}$}~\cite{xenon_xe1t-sensitivity}.

In the planned upgrade of XENON1T to XENONnT, the LXe target mass will increase to a total of $\sim$6 tons. This will require a \mbox{$\sim$40$\%$} increase in the linear dimensions of the TPC and nearly double the number of PMTs. The larger detector will improve the sensitivity by another order of magnitude, reaching \mbox{$\sigma_\mathrm{SI} \lesssim 10^{-48}~\mathrm{cm}^{2}$} at \mbox{m$_{\chi}$= 50 GeV/c $^{2}$}~\cite{xenon_xe1t-sensitivity}, assuming a negligible contribution from materials and a total exposure of 20 ton$\cdot$years.

Most of the existing sub-systems for XENON1T were designed to be reused for XENONnT, however the upgrade requires the construction of a new TPC and inner cryostat. As material-induced ER backgrounds are expected to be even lower than in XENON1T, the screening effort and material selection is focused on reducing the nuclear recoil background. This is being addressed particularly through continued efforts to identify low-activity stainless steel and by pursuing viable alternatives to PTFE, where possible.

\section{Acknowledgments}

We gratefully acknowledge support from the National Science Foundation, Swiss National Science Foundation, Deutsche Forschungsgemeinschaft, Max Planck Gesellschaft, German Ministry for Education and Research, Netherlands Organisation for Scientific Research (NWO), Weizmann Institute of Science, I-CORE, Initial Training Network Invisibles (Marie Curie Actions, PITNGA-2011-289442), Fundacao para a Ciencia e a Tecnologia, Region des Pays de la Loire, Knut and Alice Wallenberg Foundation, Kavli Foundation, and Istituto Nazionale di Fisica Nucleare. We are grateful to Laboratori Nazionali del Gran Sasso for hosting and supporting the XENON project.


\clearpage
\thispagestyle{empty}
\newgeometry{margin=3cm}

\begin{sidewaystable*}[tp]
\resizebox{\textwidth}{!}{
\tabcolsep=4pt
\centering
\begin{tabular}{ cllllccccccccccccc}

\toprule
\\
\textbf{Item} &  \textbf{Sample}  & \textbf{Supplier} &  \textbf{XENON1T Use}  &  \textbf{Facility}  & \textbf{Mass [kg]}  &  \textbf{Time [d]}  &  \textbf{Units}  &  \textbf{$^{235}$U}  &  \textbf{$^{238}$U}  &  \textbf{$^{226}$Ra}  &  \textbf{$^{228}$Ra ($^{232}$Th)$^{\dagger}$}  &  \textbf{$^{228}$Th}  &  \textbf{$^{40}$K}  &  \textbf{$^{60}$Co}  &  \textbf{$^{137}$Cs} \\
\hline
\hline
\\
\multicolumn{16}{l}{\textbf{Copper}} \\
\hline
1 & 
Cu-OFE (CW009A) & 
Luvata & 
Field cage & 
GeMPI II & 
58.0 & 
24.8 & 
mBq/kg & 
$<0.18$ & 
$<1$ & 
$<0.035$ & 
$<0.033$ & 
$<0.026$ & 
$0.4(1)$ & 
$<0.019$ & 
$<0.033$ 
\\
1 & 
Cu-OFE (CW009A) & 
Luvata & 
Field cage & 
ICP-MS & 
$-$ & 
$-$ & 
mBq/kg & 
$-$ & 
$0.03(1)  $ & 
$-$ & 
$<0.01^{\dagger}$ & 
$-$ & 
$-$ & 
$-$ & 
$-$ & 
\\
2 & 
OFHC (C10100) & 
NBM & 
PMT arrays & 
GeMPI I& 
8.8 & 
37.2 & 
mBq/kg & 
$<0.12$ & 
$<1.9$ & 
$<0.13$ & 
$0.090(4)$ & 
$0.090(4)$ & 
$0.5(2)$ & 
$0.07(1)$ & 
$<0.08$  
\\
3 & 
OFHC (C10100) & 
NBM & 
PMT arrays & 
GeMPI II& 
23.3 & 
23.5 & 
mBq/kg & 
$<0.19$ & 
$<2.0$ & 
$<0.054$ & 
$<0.07$ & 
$<0.055$ & 
$<0.46$ & 
$0.17(2)$ & 
$<0.03$  
\\
4 & 
OFHC (C10100) & 
NBM & 
Field cage & 
GeMPI I& 
31.3 & 
42.1 & 
mBq/kg & 
$<0.15$ & 
$<2.4$ & 
$<0.073$ & 
0.10(4)& 
0.16(4) & 
$< 0.34$ & 
0.21(2) & 
$< 0.03$  
 \\
4 & 
OFHC (C10100) & 
NBM & 
Field cage & 
ICP-MS & 
$-$ & 
$-$ & 
mBq/kg & 
$-$ & 
$< 0.03  $ & 
$-$ & 
$<0.01^{\dagger}$ & 
$-$ & 
$-$ & 
$-$ & 
$-$ & 
\\
5 & 
OFHC (C10100) & 
Forster Tool & 
PMT arrays& 
ICP-MS & 
$-$ & 
$-$ & 
mBq/kg & 
$-$ & 
$< 0.03  $ & 
$-$ & 
$<0.01^{\dagger}$ & 
 $-$ &
$-$ &
$-$ &
$-$ &  
\\
6 & 
Au-plated Cu rods & 
D\"{o}rrer & 
$-$ & 
Gator & 
2.4 & 
28 & 
mBq/kg & 
$<0.94$ & 
$<20.4^{*}$ & 
$<1.1$ & 
$<1.6$ & 
$<1.1$ & 
$83(9)$ & 
$<0.49$ & 
$<0.30$  
\\ 
7 & 
Au-plated Cu, welded & 
D\"{o}rrer & 
$-$ & 
GeMPI I& 
2.2 & 
15 & 
mBq/kg & 
$<1.1$ & 
$<32$ & 
$2.0(3)$ & 
$<1.4$ & 
$<0.52$ & 
$140(20)$ & 
$0.3(2)$ & 
$<0.37$  
\\
\\
\multicolumn{16}{l}{\textbf{Stainless steel}} \\
\hline
8 & 
AISI 316Ti, 50 mm & 
NIRONIT & 
Cryostat domes & 
Gator & 
10.8 & 
7.3 & 
mBq/kg & 
$<3.9$ & 
$<150$ & 
$<4.0$ & 
$<4.8$ & 
$4.5(6)$ & 
$<5.6$ & 
$37.3(9)$ & 
$<1.5$  
\\ 
8 & 
AISI 316Ti, 50 mm & 
NIRONIT & 
Cryostat domes & 
ICP-MS & 
$-$ & 
$-$ & 
mBq/kg & 
$-$ & 
$1.5(5)$ & 
$-$ & 
$0.21(6)^{\dagger}$ & 
$-$ & 
$-$ & 
$-$ & 
$-$ & 
\\ 
8 & 
AISI 316Ti activated  & 
NIRONIT & 
Cryostat domes & 
GeMPI I& 
10.8 & 
6.9 & 
mBq/kg & 
$<4.2$ & 
$<96$ & 
$<2.2$ & 
$<5.4$ & 
$2.9(6)$ & 
$5(2)$ & 
$8.2(6)$ & 
$<0.86$  
\\ 
9 & 
AISI 304L, 5 mm  & 
NIRONIT & 
Cryostat & 
GeMPI II & 
10.1 & 
6.7 & 
mBq/kg & 
$<2.0$ & 
$<40$ & 
$<0.64$ & 
$<0.81$ & 
$<0.36$ & 
$<2.7$ & 
$9.7(8)$ & 
$<0.64$  
\\ 
9 & 
AISI 304L, 5 mm  & 
NIRONIT & 
Cryostat & 
ICP-MS & 
$-$ & 
$-$ & 
mBq/kg & 
$-$ & 
$2.4(7)$ & 
$-$ & 
$0.21(6)^{\dagger}$ & 
$-$ & 
$-$ & 
$-$ & 
$-$ & 
\\ 
10 & 
AISI 304, 5 mm & 
NIRONIT & 
Bell, Field cage & 
GeMPI II & 
10.3 & 
5.9 & 
mBq/kg & 
$<1.4$ & 
$<37$ & 
$1.2(3)$ & 
$2.1 (7)$ & 
$2.0(4)$ & 
$<1.3$ & 
$5.5(5)$ & 
$<0.58$  
\\
10 & 
AISI 304, 5 mm & 
NIRONIT & 
Bell, Field cage & 
ICP-MS & 
$-$  & 
$-$  & 
mBq/kg & 
$-$ & 
$11(3)$ & 
$-$ & 
$1.2(4)^{\dagger}$ & 
$-$ & 
$-$ & 
$-$ & 
$-$ & 
\\ 
11 & 
AISI 316L, 2 mm & 
ALCA & 
Cryostat pipe & 
Gator & 
4.5 & 
8.2 & 
mBq/kg & 
$<1.7$ & 
$<38$ & 
$<2.7$ & 
$<5.5$ & 
$4.3(6)$ & 
$<6.8$ & 
$15(1)$ & 
$<0.65$  
\\
12 & 
AISI 316L, 5 mm & 
NIRONIT & 
$-$ & 
Gator & 
10.3 & 
11 & 
mBq/kg & 
$<2.7$ & 
$<83$ & 
$4.3(4)$ & 
$4 (1)$ & 
$7.0(6)$ & 
$<3.8$ & 
$29(2)$ & 
$<0.73$  
\\
13 & 
AISI 304L, 5 mm & 
Outokumpu & 
$-$ & 
GeMPI I & 
17.9 & 
4.0 & 
mBq/kg & 
$<25$ & 
$<94$ & 
$3.1(6)$ & 
$3(1)$ & 
$6(1)$ & 
$<1.6$ & 
$36(3)$ & 
$<1.1$  
\\
14 & 
AISI 316L, 5 mm & 
NIRONIT & 
$-$ & 
GeMPI II & 
10.4 & 
7.9 & 
mBq/kg & 
$<4.9$ & 
$<140$ & 
$2.3(7)$ & 
$<4.4$ & 
$14(1)$ & 
$4(2)$ & 
$142(9)$ & 
$<1.2$  
\\
15 & 
AISI 316L, 5 mm & 
Outokumpu & 
$-$ & 
GeMPI I & 
17.9 & 
8.9 & 
mBq/kg & 
$<1.6$ & 
$<61$ & 
$2.4(3)$ & 
$1.7(7)$ & 
$2.1(3)$ & 
$<2.9$ & 
$20(2)$ & 
$<0.64$  
\\
16 & 
AISI 304, 5 mm & 
Di Zio & 
$-$ & 
GeMPI I & 
10.2 & 
8.9 & 
mBq/kg & 
$<2.1$ & 
$<70$ & 
$2.8(6)$ & 
$<2.1$ & 
$<0.75$ & 
$6(2)$ & 
$25(2)$ & 
$<0.68$  
\\
17 & 
AISI 304L, 5 mm & 
Acerinox Europa & 
$-$ & 
GeMPI II & 
9.8 & 
3.9 & 
mBq/kg & 
$<3.5$ & 
$100(50)$ & 
$<1.9$ & 
$4(1)$ & 
$30(2)$ & 
$<4.1$ & 
$15(1)$ & 
$<0.49$  
\\
18 & 
AISI 316Ti, 5 mm & 
Outokumpu & 
$-$ & 
GeMPI II & 
9.9 & 
6.8 & 
mBq/kg & 
$<2.7$ & 
$<120$ & 
$<1.0$ & 
$<2.7$ & 
$<0.9$ & 
$<4.2$ & 
$17(1)$ & 
$<1.1$  
\\
18 & 
AISI 316Ti, 5 mm& 
Outokumpu & 
$-$ & 
ICP-MS & 
$-$ & 
$-$ & 
mBq/kg & 
$-$ & 
$3.0(9)$ & 
$-$ & 
$0.7(2)^{\dagger}$ & 
$-$ & 
$-$ & 
$-$ & 
$-$  
\\
19 & 
AISI 316Ti, 5 mm & 
NIRONIT & 
$-$ & 
GeMPI II & 
8.0 & 
4.9 & 
mBq/kg & 
$<4.0$ & 
$<120$ & 
$<1.2$ & 
$<3.7$ & 
$<1.8$ & 
$<6.8$ & 
$113(8)$ & 
$<1.4$  
\\
20 & 
ASME 304L, 5 mm & 
Walter Tosto  & 
$-$ & 
GeMPI I & 
10.2 & 
8.8 & 
mBq/kg & 
$<2.1$ & 
$<54$ & 
$0.6(3)$ & 
$<1.2$ & 
$0.8(3)$ & 
$<2.9$ & 
$15(1)$ & 
$<0.43$  
\\
20 & 
ASME 304L, 5 mm & 
Walter Tosto  & 
$-$ & 
ICP-MS & 
$-$ & 
$-$ & 
mBq/kg & 
$-$ & 
$2.5(7)$ & 
$-$ & 
$0.14(4)^{\dagger}$ & 
$-$ & 
$-$ & 
$-$ & 
$-$ & 
\\
20 & 
ASME 304L, 5 mm & 
Walter Tosto  & 
$-$ & 
GeMPI II & 
9.7 & 
8.1 & 
mBq/kg & 
$<15$ & 
$<51$ & 
$0.4(2)$ & 
$<1.1$ & 
$0.8(3)$ & 
$<2.5$ & 
$13.3(1)$ & 
$<0.29$  
\\
21 & 
AISI 316, 5 mm & 
Walter Tosto & 
$-$  & 
GeMPI I & 
10.0 & 
10.8 & 
mBq/kg & 
$<1.8$ & 
$<29$ & 
$<0.55$ & 
$<0.97$ & 
$<0.5$ & 
$<3.0$ & 
$10.7(8)$ & 
$<0.46$  
\\
21 & 
AISI316, 5 mm & 
Walter Tosto & 
$-$  & 
ICP-MS & 
$-$ & 
$-$ & 
mBq/kg & 
$-$ & 
$4(1)$ & 
$-$ & 
$1.2(3)^{\dagger}$ & 
$-$ & 
$-$ & 
$-$ & 
$-$ & 
\\
22 & 
AISI 316Ti, 5 mm & 
NIRONIT  & 
$-$  & 
GeMPI I & 
10.0 & 
6.9 & 
mBq/kg & 
$<2.9$ & 
$<200$ & 
$<0.54$ & 
$<5.5$ & 
$<3.8$ & 
$3(2)$ & 
$86(6)$ & 
$<1.8$  
\\
23 & 
AISI 304, 3$-$10mm & 
Di Zio & 
$-$ & 
GeMPI II & 
20.2 & 
7.1 & 
mBq/kg & 
$<1.2$ & 
$30(1)$ & 
$2.2(3)$ & 
$1.6(4)$ & 
$2.0(3)$ & 
$1.6(4)$ & 
$11.4(9)$ & 
$0.4(2)$  
\\
24 & 
AISI 304 & 
Di Zio & 
Water shield & 
Corrado & 
1.9 & 
5.9 & 
mBq/kg & 
$-$ & 
$< 269$ & 
$< 7.3$ & 
$< 7.7$ & 
$< 7.6$ & 
$< 19.1$ & 
$19(2)$ & 
$< 3.5$ & 
\\
25 & 
AISI 304L & 
Stalatube & 
Support structure & 
GeMPI I & 
5.9 & 
50 & 
mBq/kg & 
$< 3.6$ & 
$< 59$ & 
$5.0(9)$ & 
$3.0(1)$ & 
$27(2)$ & 
$< 7.8$ & 
$4.0(6)$ & 
$< 0.43$ & 
\\
25 & 
AISI 304L & 
Stalatube & 
Support structure & 
Corrado & 
5.9 & 
10.9 & 
mBq/kg & 
$-$ & 
$< 64$ & 
$2.2(6)$ & 
$< 3.1$ & 
$10(1)$ & 
$< 5.2$ & 
$1.9(2)$ & 
$< 0.6$ & 
\\
26 & 
AISI 304L & 
Stalatube & 
Support structure & 
Corrado & 
5.7 & 
11.1 & 
mBq/kg & 
$-$ & 
$< 200$ & 
$3(1)$ & 
$< 3.8$ & 
$< 2.5$ & 
$< 12.9$ & 
$34.0(6)$ & 
$< 0.8$ & 
\\
27 & 
AISI 304L & 
Stalatube & 
Support structure & 
Corrado & 
5.7 & 
17.8 & 
mBq/kg & 
$-$ & 
$< 135$ & 
$2.6(1)$ & 
$< 2.7$ & 
$< 2.6$ & 
$< 11.9$ & 
$33(1)$ & 
$< 0.6$ & 
\\
28 & 
ASTM 304 fasteners & 
UC Components & 
Field cage & 
GeDSG & 
11.4 g & 
23.6 & 
mBq/kg & 
$< 4.2$ & 
$< 56$ & 
37(6) & 
$< 18$ & 
$< 24$ & 
$< 87$ & 
22(4) & 
$< 3.3$ & 
\\
29 & 
316L Fasteners & 
Socket Source & 
Field cage & 
GeDSG & 
8.5 g & 
14.2 & 
mBq/kg & 
$< 7$ & 
$< 110$ & 
0.15(2) & 
$< 34$ & 
$< 48$ & 
$< 160$ & 
14(6) & 
$< 9$ & 
\\
30 & 
304L Tube & 
McMaster Carr & 
$-$ & 
Gator & 
5.8 & 
19 & 
mBq/kg & 
$1.4(4)$ & 
$30.4(9)^{*}$ & 
$1.4(4)$ & 
$<2.8$ & 
$5.9(5)$ & 
$<4.9$ & 
$3.2(3)$ & 
$<0.66$ & 
\\
31 & 
304L Tube & 
McMaster Carr & 
$-$ & 
GDMS & 
$-$ & 
$-$ & 
mBq/kg & 
$-$ & 
$2.5(7)$ & 
$-$ & 
$0.8(2)^{\dagger}$ & 
$-$ & 
$-$ & 
$-$ & 
$-$ & 
\\
\\
\multicolumn{16}{l}{\textbf{Titanium}} \\
\hline
32 & 
Ti grade 1 & 
Uniti & 
$-$ & 
Gator & 
14.0 & 
6 & 
mBq/kg & 
$2.4(4)$ & 
$87(13)$ & 
$<1.3$ & 
$< 1.4$ & 
$< 1.7$ & 
$< 3.6$ & 
$< 0.30$ & 
$< 0.54$ & 
\\
33 & 
Ti grade 1 & 
Thyssen & 
$-$ & 
Gator & 
28.3 & 
13 & 
mBq/kg & 
$1.4(3)$ & 
$27(6)$ & 
$<0.58$ & 
$< 0.81$ & 
$0.7(3)$ & 
$< 1.5$ & 
$< 0.17$ & 
$< 0.24$ & 
\\
34 & 
Ti grade 4 & 
NIRONIT & 
$-$ & 
Gator & 
11.3 & 
13.5 & 
mBq/kg & 
$< 2.2$ & 
$< 42$ & 
$< 1.2$ & 
$2.1(6)$ & 
$9(1)$ & 
$< 4.9$ & 
$< 0.29$ & 
$< 0.36$ & 
\\
35 & 
Ti grade 2 & 
NIRONIT & 
$-$ & 
Gator & 
10.8 & 
17 & 
mBq/kg & 
$1.4(4)$ & 
$40(10)$ & 
$< 0.81$ & 
$1.9(6)$ & 
$3.1(3)$ & 
$< 2.9$ & 
$< 0.25$ & 
$< 0.46$ & 
\\
36 & 
Ti grade 1 & 
NIRONIT & 
$-$ & 
ICP-MS & 
$-$ & 
$-$ & 
mBq/kg & 
$-$ & 
$25(4)$ & 
$-$ & 
$< 8.1^{\dagger}$ & 
$-$ & 
$-$ & 
$-$ & 
$-$ & 
\\
37 & 
Ti grade 1 & 
NIRONIT & 
$-$ & 
GDMS & 
$-$ & 
$-$ & 
mBq/kg & 
$-$ & 
$28.0(1)$ & 
$-$ & 
$5.20(5)^{\dagger}$ & 
$-$ & 
$-$ & 
$-$ & 
$-$ & 
\\
38 & 
Ti grade 1 & 
NIRONIT & 
$-$ & 
Gator & 
10.5 & 
27.6 & 
mBq/kg & 
$1.3(3)$ & 
$30(10)$ & 
$1.1(4)$ & 
$< 1.1$ & 
$< 0.71$ & 
$< 2.8$ & 
$< 0.18$ & 
$< 0.19$ & 
\\
39 & 
Ti grade 1 ($\#$38, welded) & 
NIRONIT & 
$-$ & 
Gator & 
3.7 & 
31 & 
mBq/kg & 
$1.5(4)$ & 
$< 32$ & 
$< 1.3$ & 
$< 1.6$ & 
$0.9(4)$ & 
$< 3.4$ & 
$< 0.29$ & 
$< 0.32$ & 
\\
40 & 
Ti grade 1 & 
Supra Alloys Inc. & 
$-$ & 
Gator & 
9.8 & 
25 & 
mBq/kg & 
$3.0(8)$ & 
$90(30)$ & 
$< 0.6$ & 
$< 1.8$ & 
$0.9(2)$ & 
$< 2.5$ & 
$< 0.20$ & 
$0.2(1)$ & 
\\
\hline
\\
& &  &  &    &    &     &    &  &     &      &     &     &   & \textbf{(Continued)}           \\

\end{tabular}}
\end{sidewaystable*}


\clearpage
\thispagestyle{empty}
\newgeometry{margin=3cm}

\begin{sidewaystable*}[tp]
\resizebox{\textwidth}{!}{
\tabcolsep=4pt
\centering
\begin{tabular}{ cllllccccccccccccc}

\multicolumn{16}{l}{\textbf{(Continuation)}} \\
\toprule
\\
\textbf{Item} &  \textbf{Sample}  & \textbf{Supplier} &  \textbf{XENON1T Use}  &  \textbf{Facility}  & \textbf{Mass [kg]}  &  \textbf{Time [d]}  &  \textbf{Units}  &  \textbf{$^{235}$U}  &  \textbf{$^{238}$U}  &  \textbf{$^{226}$Ra}  &  \textbf{$^{228}$Ra ($^{232}$Th)$^{\dagger}$}  &  \textbf{$^{228}$Th}  &  \textbf{$^{40}$K}  &  \textbf{$^{60}$Co}  &  \textbf{$^{137}$Cs}  \\
\hline
\hline
\\
\multicolumn{16}{l}{\textbf{Cryostat insulation}} \\
\hline
41 & 
Mylar foil & 
Sheldahl/Multek & 
Cryostat, domes & 
ICP-MS & 
$-$ & 
$-$ & 
mBq/kg & 
$-$ & 
50(10) & 
$-$ & 
$1.6(5)^{\dagger}$ & 
$-$ & 
$-$ & 
$-$ & 
$-$  
\\ 
42 & 
Mylar foil & 
RUAG Space GmbH & 
$-$ & 
ICP-MS & 
$-$ & 
$-$ & 
mBq/kg & 
$-$ & 
$17(6)$ & 
$-$ & 
$2.8(9)^{\dagger}$ & 
$-$ & 
$-$ & 
$-$ & 
$-$  
\\ 
43 & 
Mylar foil & 
RUAG Space GmbH & 
$-$ & 
ICP-MS & 
$-$ & 
$-$ & 
mBq/kg & 
$-$ & 
$210(60)$ & 
$-$ & 
$11(3)^{\dagger}$ & 
$-$ & 
$-$ & 
$-$ & 
$-$  
\\ 
44 & 
Mylar (polyester layer) & 
RUAG Space GmbH & 
$-$ & 
ICP-MS & 
$-$ & 
$-$ & 
mBq/kg & 
$-$ & 
1.8 & 
$-$ & 
$0.7^{\dagger}$ & 
$-$ & 
$-$ & 
$-$ & 
$-$  
\\ 
45 & 
Mylar (Al/plastic layer) & 
RUAG Space GmbH & 
$-$ & 
ICP-MS & 
$-$ & 
$-$ & 
mBq/kg & 
$-$ & 
0.7 & 
$-$ & 
$0.04^{\dagger}$ & 
$-$ & 
$-$ & 
$-$ & 
$-$  
\\
\\
\multicolumn{16}{l}{\textbf{Plastics}} \\
\hline
46 &
PTFE & 
Maagtechnic & 
Field cage & 
ICP-MS & 
$-$ & 
$-$ & 
mBq/kg & 
$-$ & 
$<0.12$ & 
$-$ & 
$<0.06^{\dagger}$ & 
$-$ & 
$-$ & 
$-$ & 
$-$ & 
\\
47 & 
PTFE & 
Amsler $\&$ Frey & 
Field cage & 
ICP-MS & 
$-$ & 
$-$ & 
mBq/kg & 
$-$ & 
$<0.12$ & 
$-$ & 
$<0.04^{\dagger}$ & 
$-$ & 
$-$ & 
$-$ & 
$-$ & 
\\
48 & 
PTFE & 
Maagtechnic & 
Field cage & 
ICP-MS & 
$-$ & 
$-$ & 
mBq/kg & 
$-$ & 
$<0.12$ & 
$-$ & 
$<0.04^{\dagger}$ & 
$-$ & 
$-$ & 
$-$ & 
$-$ & 
\\
48 & 
PTFE & 
Maagtechnic & 
Field cage & 
GeMPI I & 
16.7 & 
26.4 & 
mBq/kg & 
$< 0.08$ & 
$< 1.74^{*}$ & 
$< 0.053$ & 
$< 0.060$ & 
$< 0.063$ & 
$< 0.370$ & 
$-$ & 
$< 0.024$ & 
\\  
49 & 
PTFE & 
Prof.~Plastics & 
PMT array & 
GeMPI II & 
13.1 & 
58.1 & 
mBq/kg & 
$< 0.055$ & 
$< 1.194^{*}$ & 
$0.07(2)$ & 
$< 0.071$ & 
$0.06(2)$ & 
$5.2(6)$ & 
$-$ & 
$<0.024$  
\\  
49 & 
PTFE & 
Prof.~Plastics & 
PMT array & 
ICP-MS & 
$-$ & 
$-$ & 
mBq/kg & 
$-$ & 
$0.09(1)$ & 
$-$ & 
$0.049(8)^{\dagger}$ & 
$-$ & 
$-$ & 
$-$ & 
$-$  
\\  
50 & 
PTFE & 
Amsler $\&$ Frey & 
Field cage & 
Gator & 
44 & 
58 & 
mBq/kg & 
$<0.087$ & 
$<1.96$ & 
$<0.12$ & 
$<0.11$ & 
$<0.065$ & 
$<0.343$ & 
$<0.027$ & 
$0.17(3)$  
\\  
50 & 
PTFE & 
Amsler $\&$ Frey & 
Field cage & 
ICP-MS & 
$-$ & 
$-$ & 
mBq/kg & 
$-$ & 
$<0.187$ & 
$-$ & 
$<0.041^{\dagger}$ & 
$-$ & 
$-$ & 
$-$ & 
$-$ & 
\\  
50 & 
PTFE & 
Amsler $\&$ Frey & 
Field cage & 
ICP-MS & 
$-$ & 
$-$ & 
mBq/kg & 
$-$ & 
$<0.25$ & 
$-$ & 
$<0.041^{\dagger}$ & 
$-$ & 
$-$ & 
$-$ & 
$-$ & 
\\  
51 & 
PTFE, 15$\%$ quartz & 
Unknown & 
$-$ & 
Corrado & 
0.53 & 
14.4 & 
mBq/kg & 
$-$ & 
$4000(1000)$ & 
$5000(200)$ & 
$8900(400)$ & 
$2600(200)$ & 
$4700(300)$ & 
$<12.3$ & 
$<6.0$  
\\  
52 & 
PEEK fasteners &
Solidspot & 
Field cage &
GeMPI II & 
25 g & 
20.9 & 
mBq/kg & 
$< 2.6$ & 
$<56.4^{*}$ & 
13(3) & 
$< 20$ & 
$< 10$ & 
50(30) & 
$< 1.6$ & 
6(2) & 
\\  
53 & 
PAI (Torlon 4203) & 
Cellpack & 
Field cage & 
GeMPI I& 
0.75 & 
21.6 & 
mBq/kg & 
$2.9(8)$ & 
$<41$ & 
$2.0(5)$ & 
$2.9(8)$ & 
$2.4(6)$ & 
$22(5)$ & 
$-$ & 
$< 0.46$  
\\  
54 & 
PAI (Torlon 4203L) & 
Drake Plastics & 
Field cage & 
Gator & 
1.2 & 
20 & 
mBq/kg & 
$<26$ & 
$<114$ & 
$<2.6$ & 
$<5.5$ & 
$3(1)$ & 
$11.5(1)$ & 
$<1$ & 
$<1$  
\\  
54 & 
PAI (Torlon 4203L) & 
Drake Plastics & 
Field cage & 
ICP-MS & 
$-$ & 
$-$ & 
mBq/kg & 
$-$ & 
$4.9(4)$ & 
$-$ & 
$0.49(4)^{\dagger}$ & 
$-$ & 
$-$ & 
$-$ & 
$-$  
\\
\\
\multicolumn{16}{l}{\textbf{Cables}} \\
\hline
55 & 
PTFE coax RG196 & 
koax24 & 
PMT, Field cage & 
GeMPI I& 
1.76 & 
42.5 & 
mBq/kg & 
$<1.0$ & 
$<25$ & 
$<0.59$ & 
$<0.58$ & 
$0.6(2)$ & 
$33(4)$ & 
$<0.21$ & 
$<0.10$  
\\  
56 & 
PTFE coax RG196 & 
koax24 & 
PMT, Field cage & 
GeMPI II & 
1.68 & 
12.5 & 
mBq/kg & 
$<1.5$ & 
$<27$ & 
$0.4(2)$ & 
$<0.77$ & 
$<0.56$ & 
$9(3)$ & 
$<0.27$ & 
$<0.41$  
\\  
57 & 
Kapton wire, TYP-2830 & 
Accu-Glass & 
PMT, Field cage & 
GeMPI IV & 
147 g & 
41.5 & 
mBq/kg & 
$<2.8$ & 
$<130^{*}$ & 
$4(1)$ & 
$<2.2$ & 
$<3.6$ & 
$280(30)$ & 
$<1.1$ & 
$<0.54$  
\\ 
58 & 
Kapton wire, TYP-2830 & 
Accu-Glass & 
PMT, Field cage & 
Gator & 
78 g & 
$14.2$ & 
mBq/kg & 
$<12$ & 
$<436$ & 
$<25$ & 
$<31$ & 
$<26$ & 
$460(70)$ & 
$<8.3$ & 
$<6.8$  
\\  
58 & 
Kapton wire, TYP-2830 & 
Accu-Glass & 
PMT, Field cage & 
ICP-MS & 
$-$ & 
$-$ & 
mBq/kg & 
$-$ & 
$4.2$ & 
$-$ & 
$0.71^{\dagger}$ & 
$-$ & 
$-$ & 
$-$ & 
$-$  
\\  
59 & 
PTFE coax RG196 & 
Huber-Suhner & 
$-$ & 
GeMPI II & 
0.59 & 
38.6 & 
mBq/kg & 
$<0.85$ & 
$<49$ & 
$1.0(3)$ & 
$0.7(4)$ & 
$0.8(3)$ & 
$20(4)$ & 
$<0.35$ & 
$<0.14$  
\\  
60 & 
PTFE coax RG196 & 
Huber-Suhner & 
$-$ & 
ICP-MS & 
$-$ & 
$-$ & 
mBq/kg & 
$-$ & 
$1.9(6)$ & 
$-$ & 
$2.4(7)^{\dagger}$ & 
$-$ & 
$-$ & 
$-$ & 
$-$  
\\  
61 & 
PTFE coax SM50 & 
Habia Cable & 
$-$ & 
Corrado & 
1.35 & 
11.6 & 
mBq/kg & 
$-$ & 
$< 694$ & 
10(6) & 
$< 21.6$ & 
$< 23$ & 
130(30) & 
$< 3.3$ & 
$< 6.6$ & 
\\
62 & 
Kapton coax & 
MDC Vacuum & 
$-$ & 
Gator & 
56 g & 
11.8 & 
mBq/kg & 
$<9.1$ & 
$<179$ & 
$11(3)$ & 
$<24$ & 
$<15$ & 
$800(100)$ & 
$<5.0$ & 
$<5.1$  
\\  
62 & 
Kapton coax & 
MDC Vacuum & 
$-$ & 
ICP-MS & 
$-$ & 
$-$ & 
mBq/kg & 
$-$ & 
$5.1$ & 
$-$ & 
$0.71^{\dagger}$ & 
$-$ & 
$-$ & 
$-$ & 
$-$  
\\  
63 & 
Kapton stripline & 
CERN & 
$-$ & 
Gator & 
98 g & 
13.7 & 
mBq/kg & 
$<6.1$ & 
$<226$ & 
$18(4)$ & 
$<15$ & 
$<20$ & 
$60(10)$ & 
$<4.3$ & 
$<3.3$  
\\  
63 & 
Kapton stripline & 
CERN & 
$-$ & 
ICP-MS & 
$-$ & 
$-$ & 
mBq/kg & 
$-$ & 
$100(20)$ & 
$-$ & 
$<0.4^{\dagger}$ & 
$-$ & 
$-$ & 
$-$ & 
$-$  
\\  
64 & 
Kapton stripline & 
CERN & 
$-$ & 
GePaolo & 
$-$ & 
$-$ & 
mBq/kg & 
$<22$ & 
$<560$ & 
$<33$ & 
$<40$ & 
$<12$ & 
$<390$ & 
$<4.6$ & 
$<16$  
\\
\\
\\
\multicolumn{16}{l}{\textbf{Connectors}} \\
\hline
65 & 
CuBe D-sub pins & 
Accu-Glass & 
PMT, Field cage & 
Gator & 
0.33 g & 
15 & 
$\mu$Bq/pair & 
$<0.92$ & 
$<42.3$ & 
$<1.55$ & 
$<2.28$ & 
$<1.53$ & 
$7.5(2)$ & 
$<0.57$ & 
$<0.34$  &
\\ 
65 & 
CuBe D-sub pins & 
Accu-Glass & 
PMT, Field cage & 
ICP-MS & 
$-$ & 
$-$ & 
$\mu$Bq/pair & 
$-$ & 
$<0.79$ & 
$-$ & 
$<0.26^{\dagger}$ & 
$-$ & 
$-$ & 
$-$ & 
$-$  &
\\ 
66 & 
MMCX connectors & 
Telegartner & 
PMT & 
Gator & 
442 g & 
4.2 & 
$\mu$Bq/pair & 
$20(4)$ & 
$440(80)$ & 
$<17$ & 
$38(7)$ & 
$42(5)$ & 
$180(30)$ & 
$<3.8$ & 
$<3.3$  &
\\
66  & 
MMCX connectors & 
Telegartner & 
PMT & 
ICP-MS & 
$-$ & 
$-$ & 
$\mu$Bq/pair & 
$-$ & 
$500(200)$ & 
$-$ & 
$25(8)^{\dagger}$ & 
$-$ & 
$-$ & 
$-$ & 
$-$  &
\\ 
67 & 
Au-plated Cu-alloy pins & 
ULTEMate & 
PMT, Field cage & 
GeDSG & 
35 g & 
14 & 
$\mu$Bq/pair & 
$<7.6$ & 
$<95$ & 
$18(3)$ & 
$<11.6$ & 
$<5.4$ & 
$200(50)$ & 
$<3.6$ & 
$<4.4$  &
\\ 
68 & 
CuBe D-sub pins & 
FCT electronics & 
PMT, Field cage & 
Gator & 
357 g & 
13.6 & 
$\mu$Bq/pair & 
$<0.74$ & 
$<47$ & 
$<1.1$ & 
$<1.6$ & 
$<1.7$ & 
$24(4)$ & 
$<0.46$ & 
$<0.33$  &
\\ 
68 & 
CuBe D-sub pins & 
FCT electronics & 
PMT, Field cage & 
ICP-MS & 
$-$ & 
$-$ & 
$\mu$Bq/pair & 
$-$ & 
$<2.2$ & 
$-$ & 
$<0.71^{\dagger}$ & 
$-$ & 
$-$ & 
$-$ & 
$-$ & 
\\
\hline
\\
& &  &  &    &    &     &    &  &     &      &     &     &   & \textbf{(Continued)}           \\

\end{tabular}}
\end{sidewaystable*}


\clearpage
\thispagestyle{empty}
\newgeometry{margin=3cm}

\begin{sidewaystable*}[tp]
\resizebox{\textwidth}{!}{
\tabcolsep=4pt
\centering
\begin{tabular}{ cllllccccccccccccc}

\multicolumn{16}{l}{\textbf{(Continuation)}} \\
\toprule
\\
\textbf{Item} &  \textbf{Sample}  & \textbf{Supplier} &  \textbf{XENON1T Use}  &  \textbf{Facility}  & \textbf{Mass [kg]}  &  \textbf{Time [d]}  &  \textbf{Units}  &  \textbf{$^{235}$U}  &  \textbf{$^{238}$U}  &  \textbf{$^{226}$Ra}  &  \textbf{$^{228}$Ra ($^{232}$Th)$^{\dagger}$}  &  \textbf{$^{228}$Th}  &  \textbf{$^{40}$K}  &  \textbf{$^{60}$Co}  &  \textbf{$^{137}$Cs}  \\
\hline
\hline
\\
\multicolumn{16}{l}{\textbf{Photosensors (weighted average)}} \\
\hline
69 & 
R11410-21 PMTs (332) & 
Hamamatsu & 
PMTs & 
Gator & 
194 g & 
18 d (ave) & 
mBq/PMT & 
$0.37(9)$ & 
$8(2)^{*}$ & 
$0.6(1)$ & 
$0.7(2)$ & 
$0.6(1)$ & 
$12(2)$ & 
$0.84(9)$ & 
$-$ & 
\\
70 & 
R11410-21 PMTs (40) & 
Hamamatsu & 
PMTs & 
GeMPI I& 
194 g & 
15.1 d (ave) & 
mBq/PMT & 
$-$ & 
$-$ & 
$1.0(3)$ & 
$0.5(2)$ & 
$0.6(2)$ & 
$15(3)$ & 
$1.0(12)$ & 
$-$ & 
\\
\\
\multicolumn{16}{l}{\textbf{Electronics}} \\
\hline
\\
71 & 
SMD Resistors, 5G$\mathrm{\Omega}$& 
OHMITE & 
Field cage & 
GeMi & 
11.1 g & 
13.6 & 
$\mu$Bq/piece & 
$3(1)$ & 
$120(20)$ & 
$18(2)$ & 
$6(2)$ & 
$7(1)$ & 
$50(20)$ & 
$<0.56$ & 
$<0.56$ & 
\\
72 & 
SMD Resistors & 
Vishay & 
$-$ & 
GeDSG & 
3.5 g & 
5.1 & 
$\mu$Bq/piece & 
$16(3)$ & 
$600(200)$ & 
$240(10)$ & 
$44(8)$ & 
$49(5)$ & 
$110(30)$ & 
$<5.9$ & 
$<4.2$ & 
\\
73 & 
SMD Resistors & 
OHMITE & 
$-$ & 
GeDSG & 
$-$ & 
18.5 & 
$\mu$Bq/piece & 
$9(1)$ & 
$200(20)^{*}$ & 
$25(2)$ & 
$8(2)$ & 
$9(1)$ & 
$120(20)$ & 
$<1.8$ & 
$<1.5$ & 
\\
74 & 
SMD Resistors, 15M$\mathrm{\Omega}$ & 
Stackpole Electronics & 
$-$ & 
GeCris & 
22.8 g & 
6 & 
$\mu$Bq/piece & 
$1.1(4)$ & 
$19(5)$ & 
$3.7(2)$ & 
$1.3(2)$ & 
$1.3(1)$ & 
$5.3(9)$ & 
$< 0.03$ & 
$< 0.04$ & 
\\
75 & 
SMD Resistors, 10M$\mathrm{\Omega}$ & 
Vishay & 
$-$ & 
GsOr & 
19.1 g & 
6 & 
$\mu$Bq/piece & 
$0.7(2)$ & 
$< 14$ & 
$1.53(15)$ & 
$1.0(2)$ & 
$0.74(14)$ & 
$15(2)$ & 
$<0.05$ & 
$< 0.18$ & 
\\
76 & 
SMD Resistors, 51$\mathrm{\Omega}$ & 
Vishay & 
$-$ & 
GsOr & 
21.0 g & 
20.7 & 
$\mu$Bq/piece & 
$0.6(2)$ & 
$15(3)$ & 
$1.8(1)$ & 
$1.3(1)$ & 
$1.0(1)$ & 
$17(2)$ & 
$< 0.04$ & 
$< 0.05$ & 
\\
77 & 
SMD Resistors, 1k$\mathrm{\Omega}$ & 
Vishay & 
$-$ & 
GeDSG & 
20.2 g & 
13.7 & 
$\mu$Bq/piece & 
$0.43(5)$ & 
$15(3)$ & 
$1.4(8)$ & 
$0.7(1)$ & 
$0.5(1)$ & 
$16(2)$ & 
$< 0.07$ & 
$< 0.04$ & 
\\
78 & 
SMD Resistors, 7.5 M$\mathrm{\Omega}$ & 
Rohm Semiconductor & 
$-$ & 
Corrado & 
$-$ & 
7.2 & 
$\mu$Bq/piece & 
$-$ & 
$< 60.8$ & 
12.1(8) & 
12(1) & 
10(1) & 
14(3) & 
$< 0.27$ & 
$< 0.38$ & 
\\
79 & 
SMD Resistors, 5.1 M$\mathrm{\Omega}$& 
Rohm Semiconductor & 
$-$ & 
Corrado & 
$-$ & 
10 & 
$\mu$Bq/piece & 
$-$ & 
$< 210$ & 
15(3) & 
20(5) & 
12(4) & 
$< 25.19$ & 
$< 1.15$ & 
$< 1.90$ & 
\\
80 & 
SMD Resistors, 51 $\mathrm{\Omega}$& 
Panasonic & 
$-$ & 
Corrado & 
$-$ & 
9.8 & 
$\mu$Bq/piece & 
$-$ & 
$< 322$ & 
14(2) & 
$< 12.1$ & 
$< 9.9$ & 
$< 14.1$ & 
$< 0.7$ & 
$< 1.1$ & 
\\
81 & 
SMD Resistors, 5 M$\mathrm{\Omega}$ & 
Vishay & 
PMTs & 
Bruno & 
$-$ & 
8.9 & 
$\mu$Bq/piece & 
$-$ & 
$< 40$ & 
2.1(5) & 
$< 2.8$ & 
$< 1.1$ & 
21(5) & 
$< 0.5$ & 
$< 0.8$ & 
\\
82 & 
SMD Resistors, 7.5 M$\mathrm{\Omega}$ & 
Vishay & 
PMTs & 
Gator & 
5 g & 
13.5 & 
$\mu$Bq/piece & 
0.044(9) & 
$1.0(2)^{*}$ & 
0.18(2) & 
0.14(3) & 
0.13(2) & 
1.2(2) & 
$< 0.03$ & 
$< 0.02$ & 
\\
83 & 
SMD Resistors, 51 $\mathrm{\Omega}$& 
Panasonic & 
PMTs & 
Bruno & 
$-$ & 
6.8 & 
$\mu$Bq/piece & 
$-$ & 
$< 120$ & 
$< 1.5$ & 
3(2) & 
$< 2.6$ & 
$< 10$ & 
$< 0.9$ & 
$< 0.8$ & 
\\
84 & 
SMD Resistors, 10 M$\mathrm{\Omega}$ & 
Rohm Semiconductor & 
PMTs & 
Corrado & 
$-$ & 
8.1 & 
$\mu$Bq/piece & 
$-$ & 
$< 40$ & 
5.4(4) & 
7.0(7) & 
5.7(6) & 
3.3(6) & 
$< 0.1$ & 
$< 0.3$ & 
\\
85 & 
SMD Resistors, 1 k$\mathrm{\Omega}$ & 
Panasonic & 
$-$ & 
Corrado & 
$-$ & 
5.7 & 
$\mu$Bq/piece & 
$-$ & 
$< 208$ & 
7(1) & 
5(2) & 
$< 6.1$ & 
$< 16.4$ & 
$< 0.7$ & 
$< 1.0$ & 
\\
86 & 
SMD Resistors, 1 k$\mathrm{\Omega}$  & 
Panasonic & 
PMTs & 
Corrado & 
$-$ & 
7.6 & 
$\mu$Bq/piece & 
$-$ & 
$< 145$ & 
$3(1)$ & 
$< 3.9$ & 
$< 4.0$ & 
$< 9.4$ & 
$< 0.8$ & 
$< 1.5$ & 
\\
87 & 
PEN Capacitors, 4.7 nF & 
Panasonic & 
$-$ & 
Gator & 
12.7 g & 
10 & 
$\mu$Bq/piece & 
$< 1.6$ & 
$< 69$ & 
$< 3.0$ & 
$< 3.6$ & 
7(2) & 
$< 11$ & 
$< 0.86$ & 
$< 0.82$ & 
\\
88 & 
SMD Capacitors, 10 nF & 
Unknown & 
PMTs & 
Corrado & 
$-$ & 
7.7 & 
$\mu$Bq/piece & 
$-$ & 
$< 105$ & 
63(3) & 
26(3) & 
6(2) & 
$< 13.7$ & 
$< 0.4$ & 
$< 1.2$ & 
\\
89 & 
SMD Capacitors, 10 nF & 
Kemet & 
$-$ & 
Corrado & 
61 g & 
12.5 & 
$\mu$Bq/piece & 
$-$ & 
$< 390$ & 
9(3) & 
$< 13.4$ & 
14(5) & 
60(20) & 
$< 2.7$ & 
$< 3.6$ & 
\\
90 & 
SMD Capacitors, 8.4 nF & 
AVX & 
$-$ & 
Corrado & 
140 g & 
6.9 & 
$\mu$Bq/piece & 
$-$ & 
$< 3370$ & 
4800(100) & 
780(60) & 
730(50) & 
2100(200) & 
$< 12.0$ & 
$< 6.7$ & 
\\
91 & 
SMD Capacitors, 4.7 nF & 
Panasonic & 
$-$ & 
Bruno & 
18 g & 
7.1 & 
$\mu$Bq/piece & 
$-$ & 
$< 150$ & 
$4(2)$ & 
$< 5.9$ & 
$< 3.5$ & 
$< 29$ & 
$< 1.5$ & 
$< 2.1$ & 
\\
92 & 
SMD Capacitors 10 nF & 
Kemet & 
$-$ & 
Bruno & 
61 g & 
7.9 & 
$\mu$Bq/piece & 
$-$ & 
$< 530$ & 
11(6) & 
$< 22.5$ & 
$< 16.3$ & 
$< 55$ & 
$< 5$ & 
$< 6$ & 
\\
93 & 
Sockets A24875-ND & 
TE Connectivity & 
PMTs & 
GeMPI I& 
$-$ & 
18.8 & 
$\mu$Bq/piece & 
3.7(4) & 
100(20) & 
0.26(9) & 
0.9(2) & 
1.5(2) & 
3.1(8) & 
$< 0.08$ & 
$< 0.09$ & 
\\
94 & 
Cirlex PCB & 
Fralock & 
PMTs & 
GeMPI II & 
185.1 g & 
18.8 & 
$\mu$Bq/piece & 
$< 3.0$ & 
$< 160$ & 
15(2) & 
$< 4.6$ & 
3(1) & 
33(10) & 
$< 0.98$ & 
$< 1.23$ & 
\\
95 & 
Ferrite PCB & 
Nikhef & 
$-$ & 
Corrado & 
$-$ & 
6.2 & 
mBq/piece & 
$-$ & 
$< 145$ & 
62(3) & 
53(4) & 
56(4) & 
22(6) & 
$< 0.9$ & 
$-$ & 
\\
96 & 
Ferrite PCB & 
Nikhef & 
$-$ & 
Bruno & 
$-$ & 
2.3 & 
mBq/piece & 
$-$ & 
$< 44$ & 
11.3(9) & 
$< 2.8$ & 
4.3(9) & 
$< 7.0$ & 
$< 0.3$ & 
$-$ & 
\\
97 & 
Cuflon PCB & 
Polyflon & 
$-$ & 
Bruno & 
49 g & 
8.9 & 
mBq/piece & 
$-$ & 
$< 6.43$ & 
$< 0.22$ & 
$< 0.51$ & 
$< 0.33$ & 
$< 1.9$ & 
$< 0.1$ & 
$< 0.1$ & 
\\
98 & 
Cuflon PCB & 
Polyflon & 
$-$ & 
GeMPI I& 
961 g & 
4.3 & 
mBq/piece & 
$< 3.6$ & 
$< 125$ & 
$< 0.82$ & 
$< 2.9$ & 
$< 1.9$ & 
50(10) & 
$< 3.1$ & 
$< 1.6$ & 
\\
99 & 
Solder & 
Microbond & 
$-$ & 
GeDSG & 
8.8 g & 
30.5 & 
mBq/kg & 
$< 6.7$ & 
$< 145.5^{*}$ & 
$12(5)$ & 
$< 12$ & 
$< 20$ & 
$< 0.12$ & 
$< 8.7$ & 
$< 4.9$ & 
\\
100 & 
Assembled PMT Bases & 
Various & 
PMTs & 
GeMPI I & 
230 g & 
13.5 & 
$\mu$Bq/PMT & 
66(14) & 
1100(600) & 
340(20) & 
160(20) & 
160(10) & 
280(60) & 
$< 7.5$ & 
$< 4.9$ & 
\\
\\
\multicolumn{16}{l}{\textbf{Miscellaneous}} \\
\hline
\\
101 & 
Levelmeters (SS, PEEK, PTFE)  & 
Various & 
Field Cage & 
GeMPI II & 
237.7 g & 
14.6 & 
mBq/kg & 
$< 0.62$ & 
$< 13.46^{*}$ & 
1.4(7) & 
$< 4.7$ & 
4(1) & 
26(9) & 
17(3) & 
$< 0.59$ & 
\\
102 & 
Levelmeters (Cu, PEEK) & 
Various & 
Field Cage & 
GeMPI II & 
166.8 g & 
14 & 
mBq/kg & 
$< 0.67$ & 
$< 14.55^{*}$ & 
$< 1.1$ & 
$< 1.2$ & 
$< 1.3$ & 
22(8) & 
$< 0.56$ & 
$< 0.61$ & 
\\
103 & 
Conductors (PTFE, Cu) & 
Various & 
Field Cage & 
GeMPI IV & 
61.1 g & 
39 & 
mBq/kg & 
$< 1.1$ & 
$< 23.9^{*}$ & 
3.9(8) & 
$< 1.1$ & 
$< 1.8$ & 
20(10) & 
$< 1.8$ & 
$< 0.84$ & 
\\
104 & 
SS/Polyurethane Belt & 
BRECOflex & 
Calibration system & 
Corrado & 
0.9 & 
14 & 
mBq/kg & 
$-$ & 
$<716$ & 
$10(5)$ & 
$<15.3$ & 
$<17$ & 
$<57.4$ & 
$<3.1$ & 
$<3.1$ & 
\\
104 & 
SS/Polyurethane Belt & 
BRECOflex & 
Calibration system & 
GeMPI IV & 
0.6 & 
9.5 & 
mBq/kg & 
$<2.5$ & 
$<100$ & 
$13(1)$ & 
$5(1)$ & 
$4.(1)$ & 
$90(20)$ & 
$1.9(6)$ & 
$<0.72$ & 
\\
105 & 
Feedthrough, HTBLM-133-1 & 
MDC Vacuum & 
Level control & 
Giove & 
0.7 & 
4.7 & 
mBq/kg & 
$-$ & 
$<323$ & 
$<4.6$ & 
$<13.6$ & 
$7(3)$ & 
$<14.5$ & 
$29(3)$ & 
$<4.3$ & 
\\
106 & 
Feedthrough, BRLM-133 & 
MDC Vacuum  & 
$-$ & 
Giove & 
0.85 & 
5.8 & 
mBq/kg & 
$-$ & 
$6100(600)$ & 
$9(4)$ & 
$28(6)$ & 
$510(30)$ & 
$46(1)$ & 
$8(2)$ & 
$<3.1$ & 
\\
107 & 
SS flange/Viton O-ring & 
Various & 
Level control & 
Corrado & 
1.08 & 
7 & 
mBq/kg & 
$-$ & 
$300(200)$ & 
$<4.7$ & 
$<13.4$ & 
$<6.2$ & 
$<34.3$ & 
$3(1)$ & 
$<3.9$ & 
\\
108 & 
UHMWP (polyethylene)& 
McMaster Carr & 
High-voltage feedthrough & 
Gator & 
3.83 & 
25 & 
mBq/kg & 
$<1.3$ & 
$<28$ & 
$0.8(3)$ & 
$<1.4$ & 
$<1.2$ & 
$<4$ & 
$<0.3$ & 
$<0.4$ & 
\\
\bottomrule

\end{tabular}}
\caption{\small{Measured activities of material samples for the XENON1T radioassay program. For HPGe spectrometer measurements, uncertainties are $\pm1\sigma$ for detected lines (in parentheses) and 95$\%$ C.L. for upper limits. Mass spectrometry measurement uncertainty is 30$\%$ unless otherwise noted. Listed activities for $^{238}$U that are denoted with ``*" were inferred assuming natural abundance~\cite{UraniumRatio}. Direct mass spectrometry measurements of $^{232}$Th are denoted with ``$^{\dagger}$". The ``XENON1T Use" column refers to materials that were used in the final assembly, with reference to the subgroups labeled in Figs \ref{fig:TPC_section} and \ref{fig:TPC_exploded}. Dash marks in the table indicate that the sample was not used in construction (``XENON1T Use"), or that an attribute (``Mass", ``Time", or a particular isotope) was either unknown or not applicable to the measurement.}}
\label{table:Samples}

\end{sidewaystable*}

\restoregeometry


\clearpage
\thispagestyle{empty}

\begin{table*}[ht]
\centering
\begin{tabular}{clccccccc}

\toprule
\textbf{Item} &  \textbf{Sample}  & \textbf{Units}  &  \textbf{$^{56}$Co}  &  \textbf{$^{57}$Co}  &  \textbf{$^{58}$Co}  &  \textbf{$^{54}$Mn}  &  \textbf{$^{46}$Sc} \\  
\midrule[0.06em]
1 &  Copper, CW009A  &  mBq/kg  &  0.06(2)  &  0.2(1)  &  0.36(4)  & $<$ 0.027 &  $-$  & \\
2 &  Copper, C10100 &  mBq/kg  &  0.31(3)  &  0.4(1)  &  1.8(2)  & 0.22(3) &  0.08(2)  & \\
3 &  Copper, C10100  &  mBq/kg  &  $-$  &  0.40(1)  &  0.35(4)  & 0.15(2) &  $-$  & \\
4 &  Copper,  C10100 &  mBq/kg  &  0.15(2)  &  0.7(2)  &  1.1(1)  & 0.35(4) &  $-$  & \\
8 &  Stainless steel, AISI 316Ti  &  mBq/kg  &  $<$ 0.8  &  $<$ 7.4  &  $<$ 1.5  & 1.2(5) &  $-$  & \\
9 &  Stainless steel, AISI 304L  &  mBq/kg  & $-$ & $-$ & $<$ 0.6 &  0.5(2) & $-$ & \\
10 &  Stainless steel, AISI 304  &  mBq/kg  & $-$ & $-$  & $-$  & 1.1(3) &  $-$ & \\
11 &  Stainless steel, AISI 316L  &  mBq/kg  & $-$ & $-$ & $-$ & 1.4(3) & $-$ &\\
32 &  Titanium, grade 1  &  mBq/kg  & $-$ & $-$ & $-$ & $-$ & 2.15(3) &\\
33 &  Titanium, grade 1 &  mBq/kg  & $-$ & $-$ & $-$ & $-$ & 1.9(2) &\\
34 &  Titanium, grade 4 &  mBq/kg  & $-$ & $-$ & $-$ & $-$ & 1.9(2) &\\
35 &  Titanium, grade 2  &  mBq/kg  & $-$ & $-$ & $-$ & $-$ & 2.7(3) &\\
38 &  Titanium, grade 1 &  mBq/kg  & $-$ & $-$ & $-$ & $-$ & 1.8(2) &\\
39 &  Titanium, grade 1 ($\#$38, welded) &  mBq/kg  & $-$ & $-$ & $-$ & $-$ & 1.0(1) &\\
40 &  Titanium, grade 1  &  mBq/kg  & $-$ & $-$ & $-$ & $-$ & 2.2(3) &\\
 \bottomrule
 
\end{tabular}
\caption{Cosmogenic radioisotopes detected in metal samples. The ``Item" numbers are cross-referenced with those in Table \ref{table:Samples}. Uncertainties are $\pm1\sigma$ for detected lines (in parentheses) and 95$\%$ C.L. for upper limits.}
\label{table:metalother}
\end{table*}


\clearpage
\restoregeometry
\twocolumn

\end{document}